%% file: main.tex
\documentclass[conference]{IEEEtran}

\usepackage{hyperref}
\usepackage{graphicx,url}
\usepackage{adjustbox}
\usepackage{caption,subcaption}
\usepackage[UKenglish]{babel}
\usepackage{blindtext}
\usepackage{array}
\usepackage{multirow}
\let\labelindent\relax
\usepackage{enumitem}
\usepackage{booktabs}
\usepackage{tikz}
\usepackage{soul}
\newcommand*{\eg}{\emph{e.g.,}\@\xspace}
\newcommand*{\ie}{\emph{i.e.,}\@\xspace}

\newcommand*{\etc}{\emph{etc.}\@\xspace}

\begin{document}

\title{On the Anonymity of Peer-To-Peer Network Anonymity Schemes Used by Cryptocurrencies}

\author{\IEEEauthorblockN{Piyush Kumar Sharma}
\IEEEauthorblockA{imec-COSIC, KU Leuven\\
pkumar@esat.kuleuven.be}
\and
\IEEEauthorblockN{Devashish Gosain}
\IEEEauthorblockA{Max Planck Institute for Informatics\\
dgosain@mpi-inf.mpg.de}
\and
\IEEEauthorblockN{Claudia Diaz}
\IEEEauthorblockA{imec-COSIC, KU Leuven\\
 Nym Technologies SA\\
cdiaz@esat.kuleuven.be}}


\maketitle

\pagestyle{plain}

\begin{abstract}
Cryptocurrency systems can be subject to deanonymization attacks by exploiting the network-level communication on their peer-to-peer network. Adversaries who control a set of colluding node(s) within the peer-to-peer network can observe transactions being exchanged and infer the parties involved. Thus, various network anonymity schemes have been proposed to mitigate this problem, with some solutions providing theoretical anonymity guarantees.

In this work, we model such peer-to-peer network anonymity solutions and evaluate their anonymity guarantees. To do so, we propose a novel framework that uses Bayesian inference to obtain the probability distributions linking transactions to their possible originators. We characterize transaction anonymity with those distributions, using entropy as metric of adversarial uncertainty on the originator's identity. 
In particular, we model Dandelion, Dandelion++ and Lightning Network. 
We study different configurations and demonstrate that none of them offers acceptable anonymity to their users. For instance, our analysis reveals that in the widely deployed Lightning Network, with $1\%$ strategically chosen colluding nodes, the adversary can uniquely determine the originator for $\approx50\%$ of the total transactions in the network.
In Dandelion, an adversary that controls $15\%$ of the nodes has, on average, uncertainty among only $8$ possible originators. Moreover, we observe that due to the way Dandelion and Dandelion++ are designed, increasing the network size does not correspond to an increase in the anonymity set of potential originators. Alarmingly, our longitudinal analysis of the Lightning Network reveals rather an inverse trend---with the growth of the network, the overall anonymity decreases.

\end{abstract}

\input{sections/intro}

\input{sections/Background}

\input{sections/approach_new}

\input{sections/Results}

\input{sections/Discussion}

\input{sections/Related}

\section{Conclusion}
Existing research advocates the use of anonymous P2P routing schemes to strengthen network privacy in Bitcoin and other cryptocurrency networks. Notably, hop-by-hop routing solutions (Dandelion and Dandelion++) have been proposed to anonymize the broadcast of Bitcoin transactions, while Lightning Network uses a source-routed scheme to provide payment channels as layer-2 of bitcoin. 
In this work, we propose a generic framework to measure the anonymity provided by such P2P schemes. This framework relies on Bayesian inference for computing the probability of potential originators for any given transaction and computes the overall uncertainty on the actual originator by measuring the entropy of the distribution. Our evaluation of the said schemes reveals some serious concerns. For instance, our analysis of real Lightning Network snapshots reveals that, by colluding with a few (\eg $1\%$) influential nodes, an adversary can fully deanonymize about $50\%$ of total transactions in the network. Similarly, Dandelion and Dandelion++ provide a low median entropy of $3$ and $5$ bit, respectively, when $20\%$ of the nodes are adversarial.

Overall, our study highlights a pressing need for better network anonymity schemes in cryptocurrency networks, as the current solutions provide poor anonymity to users. Our simulation framework can be used not only to evaluate existing proposals but also for designing and evaluating new solutions with a principled understanding of how much anonymity they will provide in concrete configurations. 


\bibliographystyle{IEEEtranS}
\bibliography{references}

\input{sections/appendix}

\end{document}

%% file: sections/intro.tex
\section{Introduction}
\label{sec:Intro}

Cryptocurrencies are digital currencies that are neither issued nor backed by a centralized banking or financial authority. Instead, they rely on the decentralized verification of cryptographic transactions using blockchain technology, allowing everyone to join and contribute to securing the transaction ledger~\cite{nakamoto2008bitcoin}. 
Decentralized currencies attempt to address concerns of the existing banking system, where centralization implies having entities (banks) with disproportionate power to exclude users, control financial flows, and amass a wealth of personal financial information on their customers (habits, lifestyle, spending behaviour, degree of financial desperation) that can be mined for profiling, sold and end up being used against the person. 
A growing number of blockchain-based cryptocurrencies have been proposed and deployed, with Bitcoin~\cite{Bitcoin,bohme2015bitcoin}, active for almost a decade now, being the most popular currency as well as the seminal system that popularized the concept. 

The emergence of blockchain-based cryptocurrency systems has attracted growing interest in various aspects of the underlying technologies. The decentralization and scalability aspects of blockchains have received considerable attention and are by now well understood~\cite{kim2018survey,zhou2020solutions,hafid2020scaling}. On the other hand, understanding the privacy properties of these systems presents additional complexity. Transaction anonymity requires protection both on-chain and in the underlying peer-to-peer network used to transport the transaction. 
Ideally, if a transaction is considered private, it should not be possible for third parties to identify its source or destination, neither by analyzing blockchain data nor by analyzing network traffic data available to peers. The default process of flooding transactions in the Bitcoin network has however been shown to be prone to network-level deanonymization attacks~\cite{biryukov2014deanonymisation, androulaki2013evaluating,meiklejohn2013fistful}, where the public identifier of a transaction originator (public key) is mapped to its IP address. The subsequently proposed diffusion technique (a sophisticated version of flooding) has also been shown to be vulnerable to deanonymization by adversaries that control a number of nodes in the Bitcoin network~\cite{fanti2017deanonymization, fanti2017anonymity}. 

Peer-to-peer schemes with privacy-enhanced routing features have been proposed to improve transaction anonymity in Bitcoin towards malicious peers. Dandelion~\cite{dan} and Dandelion++ \cite{dan++} have been specially designed to provide anonymity when broadcasting transactions through the Bitcoin peer-to-peer layer; whereas the Lightning Network (LN)~\cite{LN} is a layer-2 payment channel network that addresses both scalability and privacy challenges in Bitcoin. Characterizing and quantifying the anonymity provided by such peer-to-peer routing schemes, such that they can be systematically evaluated and compared, has so far remained an open challenge.

To address this challenge, this paper proposes a Bayesian framework to model and analyze the anonymity provided by peer-to-peer networking schemes toward corrupted peers. Given a network, the scheme's routing features and constraints, and an adversarial observation, the framework computes the probability distributions linking observed transactions to their possible originators. Following prior work on network anonymity metrics~\cite{diaz2002towards, Serjantov-anonymity}, we quantify the uncertainty of the adversary in determining a transaction's originator with the entropy of the probability distribution that relates the transaction to its potential originators. 

We remark that our anonymity analysis relies on network-level traffic data related to anonymously routing a transaction. Thus, the analysis is identical if originators use the peer-to-peer routing scheme to anonymously broadcast (or to send to selected recipients) a 280-character text message instead of a blockchain transaction. 
This makes the proposed Bayesian approach  applicable to evaluating \textit{anonymous peer-to-peer routing} schemes in a generic sense, regardless of whether the scheme is intended for enabling private browsing~\cite{reiter1998crowds} or messaging~\cite{drac} instead of being cryptocurrency-related. At the moment, however, practical deployments of anonymous peer-to-peer schemes relate to blockchain applications, and we focus on these for our evaluation. 

We further demonstrate the generality of our approach by applying it to schemes, Dandelion, Dandelion++ and LN, that rely on fundamentally different concepts for anonymous peer-to-peer routing. Dandelion and Dandelion++ implement \textit{hop-by-hop} probabilistic routing (that ends in \textit{broadcast}), whereas in LN transactions are \textit{source-routed} (all the way to the \textit{intended recipient}).

While our approach to computing originator probabilities is generic, we additionally take into account the specific features and constraints of each routing scheme under evaluation. Using our techniques, an adversary can perform a more nuanced analysis to identify the source of the transaction. For example, since LN is source-routed and transactions are relayed along the shortest path to the transaction counterpart, the immediate predecessor and successor of a malicious node in the transaction path provide valuable information to better estimate the originator probabilities. In \S\ref{subsec:LN_approach}, we describe in detail how this is used in the anonymity evaluation. 
Similarly, multiple adversarial nodes in Dandelion and Dandelion++ can coordinate using our proposed techniques to better identify originators and reduce transaction anonymity (ref. \S\ref{subsec:hop-by-hop}).

We implemented the general Bayesian framework as well as the scheme-specific techniques in a software simulator to evaluate transaction anonymity in LN, Dandelion and Dandelion++. The results of our analysis raise concerns about the level of protection offered in practice by these systems.
We conducted a longitudinal analysis of LN and observed that since its inception (in the year 2018), the network size has been steadily growing, but the anonymity offered has been decreasing.
For example, in the present LN topology, 
the median entropy is \textit{zero} when a few strategically selected nodes (\eg top $1\%$ degree or centrality nodes) are adversarial, meaning that the adversary can confidently identify the originator of more than half the transactions they route as an intermediary. Moreover, these few adversary nodes can intercept about $99\%$ of all transactions making the conclusion even more worrisome. 
We conducted additional experiments to further assess the anonymity offered by LN in different scenarios (\eg entropy estimations considering different transaction amounts). In all such experiments, we again observed median entropy of \textit{zero} (ref. \S\ref{subsec:source_routing_results} for details).

Furthermore, our analysis of Dandelion and Dandelion++ indicates that they do not offer high anonymity either. For example, an adversary that controls
$20\%$ of the nodes in the Bitcoin peer-to-peer network intercepts on average more than $70\%$ of transactions, and to these, Dandelion offers a median entropy of less than three bits (\ie equivalent to uncertainty among \textit{eight} possible originators per transaction). 
The same fraction of adversaries in Dandelion++ intercept roughly the same fraction of transactions, but the median entropy is about five bits (\ie equivalent to $32$ possible originators per transaction) --- thus demonstrating better anonymity in comparison to Dandelion, even if the improvement is limited. We increase network size in both Dandelion and Dandelion++ to analyze whether anonymity increases accordingly, as it would be expected given that network scaling enables larger anonymity sets~\cite{anonymity-loves-company}. We find that the anonymity offered by both Dandelion and Dandelion++ remains constant instead of increasing with network size,
implying that scaling the Bitcoin peer-to-peer network would not result in better anonymity with these schemes.

To summarize the main contributions of this work:
\begin{itemize}
    \item We propose a generic Bayesian framework to evaluate network-level anonymity in peer-to-peer networks, including both hop-by-hop and source-routed schemes. 
    
    \item Using our framework, we model and evaluate three schemes, \ie Dandelion, Dandelion++ and Lightning Network, that have been proposed and deployed to support transaction anonymity in Bitcoin. 
    
    \item We present detailed evaluation results for the aforementioned schemes and observe that they generally offer poor anonymity to transactions.
    
    \item We discuss and recommend changes that can lead to improving anonymity in these schemes.
\end{itemize}

%% file: sections/Background.tex
\section{Background}
\label{sec:back}

\subsection{Bitcoin Network}
\label{subsec:Bitcoin}

Bitcoin consists of a P2P network of nodes that communicate via TCP \cite{nakamoto2008bitcoin}.
When a Bitcoin node receives (or generates) a transaction, it further broadcasts the transaction to its neighbours within the network. The neighbours then broadcast the transaction to their neighbours and so on. 
This process is iterative and after some time the transaction information reaches all the Bitcoin nodes.

At the application layer, a Bitcoin node is identified by its public key, whereas at the network layer, it is identified by its IP address. In order to provide anonymity for the originator of the transaction, 
the originator node’s IP address and public key should remain unlinkable.
This is important because all the Bitcoin transactions generated by an originator are stored on a public blockchain in plain text along with the their public key. Notably, if the originator's public key can be linked to its IP address, the transaction would be completely deanonymized.

Prior work has demonstrated various deanoymization attacks on the Bitcoin network~\cite{biryukov2014deanonymisation, androulaki2013evaluating,meiklejohn2013fistful}. These attacks typically introduce a ``supernode'' (adversary node pretending to be a normal Bitcoin node) that connects to all Bitcoin nodes and observes the timing of transactions broadcast by other nodes. In such a situation the originator node is likely the first to be seen broadcasting the transaction. Since transactions include their originator’s public key, the supernodes are able to associate transaction public keys to their originator IP
addresses with an accuracy of up to $30\%$.
To mitigate such attacks, Bitcoin introduced an alternative approach known as \textit{diffusion}, where each node waits for a randomized amount of time (chosen independently from an exponential distribution) before broadcasting transactions to its neighbors on the Bitcoin network.
It was however later shown that this diffusion does not provide much anonymity either~\cite{fanti2017anonymity,fanti2017deanonymization}.

More recently, Dandelion~\cite{dan} and Dandelion++~\cite{dan++} have been proposed to offer theoretical anonymity guarantees to cryptocurrency networks. These schemes are already deployed on Monero~\cite{Monero} and are also under consideration for deployment in the Bitcoin network~\cite{danbip}. In addition, the Lightning Network is a payment channel network that aims to address scalability and privacy concerns of Bitcoin, including network anonymity. Lightning Network is a functional and deployed system (with about 10K active nodes) currently integrated at layer-2 with Bitcoin.
In this paper, we model and evaluate the anonymity properties offered by Dandelion, Dandelion++ and Lightning Network. We now briefly introduce these approaches.

\subsection{Dandelion}

Dandelion~\cite{dan} was designed to enhance network anonymity for Bitcoin by making it harder to link a transaction to the IP of the node that originated it. 
When a node generates a transaction in Dandelion, it does not directly diffuse it to the Bitcoin network, but instead forwards it to just one of its Bitcoin network neighbours. The neighbour node then tosses a biased coin and decides to either forward the transaction to one of its own neighbours, or to diffuse it. If it forwards the transaction to a neighbour, the process is repeated, until a node eventually diffuses the transaction. Transactions are thus forwarded a random number of hops (following a geometric distribution) before being finally diffused in the Bitcoin network. The adversary may still identify the diffuser node \cite{fanti2017anonymity}, but since this is a different node than the originator, the identity of the originator remains obfuscated. 

Dandelion thus propagates transactions in two phases: (i) stem (or \textit{anonymity}) phase, and (ii) a fluff (or \textit{diffusion}) phase.
For routing transactions in the stem phase, a privacy-subgraph graph is constructed from Bitcoin's P2P graph by selecting a subset of edges. This privacy-subgraph should ideally be a Hamiltonian circuit (a line graph) consisting of all the Bitcoin nodes.  
The fluff phase of Dandelion is simply the diffusion process of the Bitcoin network.

In Dandelion, the node that generates a transaction never diffuses it directly to the Bitcoin network, and instead always forwards it to its successor in the line graph. Probabilistic forwarding is applied from the first intermediary on: when a node receives a transaction from their predecessor, it forwards it to its successor with forwarding probability $p_f$ (where $0<p_f<1$) and diffuses it with probability $1-p_f$. 
Thus each transaction propagates over the line graph for a random number of hops before entering the fluff phase, where it is diffused in the Bitcoin network. The number of hops follows a geometric distribution with average $\frac{1}{1-p_f}$. 

\subsection{Dandelion++}

Dandelion++ \cite{dan++} builds and improves upon Dandelion. It relaxes some of the idealistic assumptions made in Dandelion that may not likely hold in practice, in particular that: (1) each node generates exactly one transaction, (2) all nodes strictly adhere to the protocol, (3) all nodes run the Dandelion protocol.
The authors of Dandelion++ further demonstrated that violations of these assumptions can lead to serious deanonymization attacks in Dandelion. 

Similar to Dandelion, Dandelion++ operates in two phases---\textit{stem} and \textit{fluff}. Unlike Dandelion however, it does not use a line as the privacy-subgraph of the stem phase; instead, it constructs a quasi $4$-regular graph where each node should have both indegree and outdegree of two (\ie two predecessors and two successors).
Thus, when a node receives a transaction from any of its predecessors, it forwards it to any one of its two successors with probability $p_f/2$, and diffuses the transaction into the Bitcoin network with probability $1-p_f$ (\ie transaction enters into fluff phase). 


\subsection{Lightning Network (LN)}
\label{Sec:LN_background}
LN~\cite{poon2016bitcoin} is a payment channel network (PCN) that was primarily proposed to address the scalability concerns of Bitcoin. In this network, two peers use the Bitcoin blockchain to open and close payment channels between them. Using these channels, peers can make payments between themselves without having to use the Bitcoin blockchain.
LN not only supports direct transactions between peers that share an open channel, but indirect as well. Users who have not established a direct payment channel can still transact through other LN participants that act as intermediaries in the PCN. These intermediaries are incentivized to route payments by a fee that they can charge for the payments they forward.

LN can be easily represented as a graph consisting of (1) nodes (users) that generate and forward transactions; (2) edges (payment channels between users); and (3) edge weights ( \emph{e.g.,} fee charged for routing a  transaction via that channel). An up-to-date snapshot of the full LN graph is maintained at each node. When some node (Alice) wants to make a payment to another node (Bob) to which it is not directly connected, Alice first computes the shortest path to Bob (using Dijkstra) in the network that charges the lowest fee (while also minimizing other factors such as the wait time in case of payment failure). 
Once Alice has computed the path, she encrypts the transaction multiple times using the Sphinx packet format~\cite{sphinx}. 
Sphinx conceals the position of the node in the path and thus Alice's successor cannot tell that it is the first node after the originator, and neither can Bob's predecessor determine that Bob is the recipient based on Sphinx packet headers.

\color{black}
There exist multiple LN client implementations, such as LND~\cite{lnd}, c-lightning~\cite{c-lightning} and eclair~\cite{eclair}, with LND being the most widely used (more than $90\%$ of clients  \cite{kumble2021lightning, rohrer2020counting, mizrahi2021congestion}). In all these implementations, senders find paths to recipients using the Dijkstra algorithm mentioned above. Each implementation uses, however, a slightly different cost function for determining edge weights in the graph, depending on factors such as the fees, timelock, and transaction amount. For example, the LND cost function is defined as:

$cost = tx\_amount*proportional\_fee + base\_fee + tx\_amount*timelock*rf + bias\_factor$

\noindent where \texttt{tx\_amount} is the transaction amount to be transferred, \texttt{proportional\_fee} is a fee that depends on the transaction amount, and \texttt{base\_fee} is a fixed fee amount for routing a payment via the channel. The \texttt{timelock} represents the amount of time required by the channel to obtain its fee in case of a successful payment, while \texttt{rf} is the risk factor accounting for the transaction amount and the time during which the amount may be unavailable in case of a payment failure (in the LND implementation the value of \texttt{rf} is in the order of $10^{-9}$). Lastly, the \texttt{bias\_factor} accounts for past payment failures via this channel, taking into account when they happened: a recent failure within the last hour introduces the largest bias, which then decreases exponentially with time as $100/(1 - \frac{1}{2^t})$, where $t$ is the number of hours elapsed since the  failure was observed~\cite{kumble2021comparative}.

Notably, LND's cost function prioritizes fees as the major criteria for selecting paths, as the timelock parameter is heavily scaled down via the risk factor, and the bias is relevant after a failure has occurred. Other implementations prioritize different aspects, with c-lightning favouring lower timelocks and eclair favouring capacity and age~\cite{kumble2021comparative}.


\color{black}

%% file: sections/approach_new.tex
\section{Threat Model and Anonymity Metric}

There are two main threat models of interest in network anonymity: \textit{global passive adversaries} and adversaries that control a subset of \textit{corrupted nodes}. 
The three schemes studied in this paper offer no anonymity protection against global passive adversaries. Such adversaries can trivially identify transaction originators using timing: nodes forward transactions shortly after receiving them, and thus whenever a node sends a transaction without having recently received one, the node is an originator; if on the other hand a node sends a transaction shortly after receiving one, then it is an intermediary. 
Protecting against such adversaries requires some notion of mixing~\cite{chaum-mix} and the introduction of per-node added latency~\cite{trilemma}. 

We thus focus on adversaries that control a subset of nodes, whether by setting up servers, hiring botnets or compromising existing nodes in the network. The adversarial nodes follow the protocols normally (the attacks are passive) but record information that they analyze with the aim to deanonymize the transactions of benign nodes (\ie identify originator nodes for each transaction). 

We consider that the adversary does not have informative priors on the activity of different nodes, and thus use uniform priors over all benign nodes (\ie we assume that \textit{a priori} all nodes are equally likely to generate a transaction). Note that a non-uniform prior informed by node activity characteristics would further reduce anonymity and facilitate transaction deanonymization. 
Starting from the uninformed prior, knowledge of the network graph and of protocol parameters, the adversary records observations from all of its nodes during operations (transactions being forwarded).
In particular, in LN, the adversary takes note of its immediate predecessor (that forwarded the transaction) as well as its immediate successor (to which the transaction is forwarded), while in Dandelion and Dandelion++, only the predecessor is relevant. 

Combining priors, known constraints and observations in a Bayesian framework, the adversary, obtains for each transaction an  \textit{a posteriori} probability distribution over all possible originators. The entropy of this distribution expresses the adversary's uncertainty about the identity of the originator. An entropy of zero means that the transaction originator can be fully and certainly identified, while an entropy of $b$ bits implies that the effective anonymity set of the transaction is equivalent to $2^b$ possible originators. 

\noindent \textbf{\textit{Rationale for using entropy:}}
Entropy metrics~\cite{diaz2002towards, Serjantov-anonymity} have been used extensively to quantify network anonymity against traffic analysis, as they allow for an intuitive interpretation of how `hidden' the true originator is among the other possible originators in the \textit{anonymity set}~\cite{anon-terminology}. This is informative of the adversarial confidence in each deanonymization (entropy zero means that the adversary is completely certain of who is the transaction originator).
The metric takes into account both the number of suspects that could be the originator, as well as how salient some suspects are with respect to others. Rather than simply assuming that the adversary guesses the most likely suspect as the originator of a transaction of interest~\cite{empirical-LN, dan}, our analysis considers  all nodes as possible originators, and computes their probability of being the true originator based on the available information.\footnote{After calculating the probabilities of each node being the originator, the adversary could also use other metrics \eg min-entropy \cite{shmatikov2006measuring} (that captures the probability of the \textit{likeliest} originator) for further analysis.} Considering the entropy of the distribution over all possible originators allows to evaluate the size and scaling of anonymity sets beyond binary successful/failed identification.

Other metrics are also possible. The Dandelion and Dandelion++ proposals~\cite{dan} model anonymity as a classification problem and evaluate anonymity using precision and recall, considering a mapping (perfect match) between transactions and originators. Based on the obtained classification accuracy, these schemes are claimed to provide `system-wide' anonymity to the transactions. In practice however, and as confirmed in our experiments, both Dandelion and Dandelion++ may offer zero anonymity to \textit{some} transactions, and thus despite `system-wide' claims, not all transactions enjoy the same level of protection. Our approach captures these differences by computing entropy per transaction and examining the distribution of anonymity for all possible transactions. 

\noindent\textbf{\textit{Assumptions about network topology:}}
Finally, we assume that the adversary has an updated view on the nodes that compose the network. 
We consider that individual nodes join and leave the P2P network as defined in the protocols of the respective scheme. In Dandelion(++) nodes can join or leave the network at any moment by connecting to another peer, while in LN nodes must create a payment channel to join the network. New nodes may not always succeed in joining the network if the existing network nodes do not accept incoming connections (Dandelion and Dandelion++) or if they do not want to open a payment channel (LN). The peer finding process in LN is straightforward as the topology is publicly available and updated periodically. In the bitcoin network however, nodes learn about other peers from their connections via gossip protocols. Moreover, node churn in these networks may impact our analysis, which we discuss in detail later in \S\ref{subsubsec:churn}.

Lastly, note that our analysis is primarily based on the network topology and the route selection policies, making abstraction of transaction details. We however account for transaction amounts in our cost analysis and we exploit transaction identifiers for easily identifying duplicate transactions received by multiple adversary nodes. Note that identifying likely duplicates may still be possible without identifiers by correlating observed transaction timings. 

 

\section{Approach}

In this section, we describe our modelling approach. Peer-to-peer anonymous routing schemes can be broadly divided into two categories---\textit{hop-by-hop} (Dandelion and Dandelion++) and \textit{source-routed} (Lightning Network). We explain the common elements of our model first and then the specifics for each type of routing.

Overall, our approach involves calculating the Bayesian (\textit{a posteriori}) probabilities of the possible transaction originators for any transaction observed by an adversary node. The entropy of this probability distribution provides a measure of transaction anonymity. We begin by introducing the basic notations and definitions that are common to both categories, followed by their detailed anonymity analysis.

\begin{itemize}
    \item $N$ is the total number of nodes in the network. 
    
    \item $C$ is the total number of adversary nodes in the network. Thus, $N-C$ is the total benign (or honest) nodes. 

    \item $B_i$ is an event that a benign node $i$ generated a transaction.
    
    \item $A_j$ is an event that an adversary node $j$ received a transaction (as an intermediary).
    
    \item $P(B_i)$ is the probability that a benign node $i$ generated a transaction. 
    
    \item $P(A_j)$ is the probability that an adversary node $j$ receives a transaction originating from any honest node.
    
    \ie the sum of probabilities of each benign node generating a transaction and forwarding it to the adversary node $j$.
    
    \item $P(A_j|B_i)$ is the conditional probability of an adversary node $j$ receiving a transaction given that honest node $i$ generated it.
    
    \item $P(B_i|A_j)$ is the conditional probability of an honest node $i$ being the originator of the transaction given that the adversary node $j$ received it.
    
\end{itemize}

Our aim is to find the probability distribution of possible originators for the transactions received by adversary nodes. These probabilities can be calculated (using Bayes' theorem) as:

\begin{equation} \label{eq:baes_thrm}
 P(B_i|A_j) = \frac{P(B_i)*P(A_j|B_i)}{P(A_j)} \ \forall i,j
\end{equation}

$P(A_j)$ can be calculated multiplying the probabilities of each benign node $i$ generating a transaction (\ie $P(B_i)$) and that transaction reaching the adversary node $j$ (\ie $P(A_j|B_i)$). This needs to be summed up for all the $N-C$ benign nodes in the network. Thus, $P(A_j)$ is computed as:

$$    P(A_j) =  \sum_{k=1}^{N-C} P(B_k) * P(A_j|B_k) $$

We consider that \textit{a priori} all benign nodes are equally likely to be the originators and thus $P(B_i) = \frac{1}{N-C}$ for all benign nodes.
Thus Eq.~(\ref{eq:baes_thrm}) can be reduced to:

\begin{equation} \label{eq:baes_thrm_reduced}
 P(B_i|A_j) = \frac{P(A_j|B_i)}{\sum_{k=1}^{N-C} P(A_j|B_k)} 
\end{equation}

Finally, we compute the anonymity of a transaction intercepted by adversary node $j$ as the Shannon entropy of the \textit{a posteriori} distribution $P(B_i|A_j)$, expressing the likelihood that honest node $B_i$ is the transaction's originator.

\begin{equation} \label{eq:Entropy_Dan}
H = - \sum_{i=1}^{N-C}P(B_i|A_j) * \log_2(P(B_i|A_j)) 
\end{equation}

We next describe scheme-specific strategies to account for routing choices and constraints when computing $P(A_j|B_i)$. 

\subsection{Hop-by-hop Routing}
\label{subsec:hop-by-hop}

Both Dandelion and Dandelion++ route  transactions using \textit{hop-by-hop} probabilistic routing, where each routing intermediary decides whether to forward to another (intermediary) hop, or to broadcast the transaction. 

\subsubsection{Dandelion}
As described earlier, Dandelion operates in two phases \ie the stem phase and the fluff phase. In Dandelion anonymity is provided \textit{only} in the stem phase (and not in the fluff phase), and the challenge is thus to identify the originator of a transaction in the stem phase. Hence we model and compute the anonymity properties of Dandelion's stem phase.

\begin{figure}[h!]
	\centering
	\includegraphics[scale=0.35]{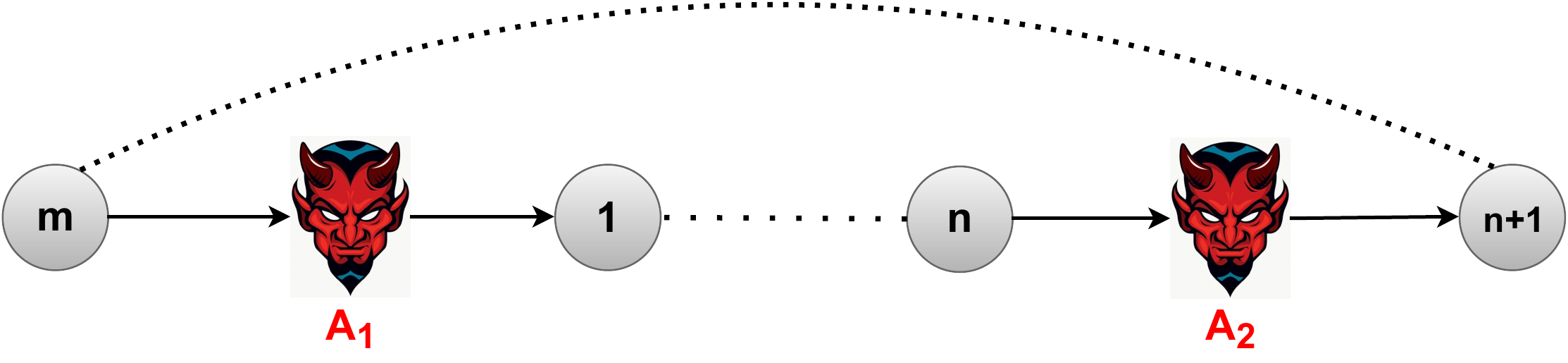}
	\caption{Dandelion privacy-subgraph: There exist $n$ benign nodes between adversary nodes $A_1$ and $A_2$; $m$ benign nodes between $A_2$ and $A_1$.}
	\label{fig:Dan_adv}
	\vspace{-2mm}
\end{figure}

In the stem phase all transactions are forwarded over a fixed line graph,
also known as the privacy-subgraph, that constrains the possible routes followed by transactions in this phase.
We assume that this privacy-subgraph is known to the adversary\footnote{We discuss in \S~\ref{subsec:grpah_learn} how the adversary can learn the privacy graph.} and model the originator probabilities for a line graph.
To do so, we first construct a line graph, and then randomly select a few nodes to be adversarial.
As a consequence, various \textbf{partitions} are created within line graph, with varying sets of benign nodes in between each pair of adversary nodes.
For example, in Fig.~\ref{fig:Dan_adv}, there are two partitions---one with benign nodes $1$ to $n$ between adversary nodes $A_1$ and $A_2$ and other with the remaining benign nodes ($n+1$ to $m$) between $A_2$ and $A_1$.
Thus, whenever an adversary node receives a transaction, it knows from which partition the transaction has come, and thus considers only the benign nodes in said partition. For instance, 
if $A_2$ receives a transaction and $A_1$ has not seen this transaction, then $A_2$ is sure that the originator is one of the nodes $1$ to $n$.
This effectively reduces the set of potential originators for any given transaction.

After learning the partition $N_P$ that contains the source of the transaction, the adversary needs to compute the conditional probabilities $P(B_i|A_j)$ for individual benign nodes in $B_i \in N_P$. 
Note that in this case, $P(A_j|B_i)=0$ for benign nodes $B_i \not \in N_P$ (Eq.~(\ref{eq:baes_thrm_reduced})). For nodes $B_i \in N_P$, $P(A_j|B_i)$ is computed as:
$$    P(A_j|B_i) = p_f^{h_{ij}-1} $$
where $h_{ij}$ is the number of hops between a benign node $B_i$ and the adversary node $A_j$.
$P(A_j|B_i)$ represents the probability of the adversary node $A_j$ receiving a transaction given that node $B_i$ generated it. In this case, $p_f$ and $h_{ij}$ both have a role in determining the probability of the transaction reaching the adversary. The farther the benign node is from the adversary (in hops), the smaller the probability of the transaction reaching the adversary.  Moreover, since the originator always forwards the transaction to its successor, the adversary receives transactions originated by its predecessor with probability $1$. 

Our model incorporates an arbitrary number of adversarial nodes. If there are three adversary nodes ($A_1$, $A_2$ and $A_3$) present in the line graph, one would have three partitions (with benign nodes between $A_1$--$A_2$, $A_2$--$A_3$ and $A_3$--$A_1$). In general, $n$ adversary nodes yield $n$ partitions of the line graph and our analysis is performed accordingly for each partition.

\subsubsection{Dandelion++}
\begin{figure}[h!]
	\centering
	\includegraphics[scale=0.6]{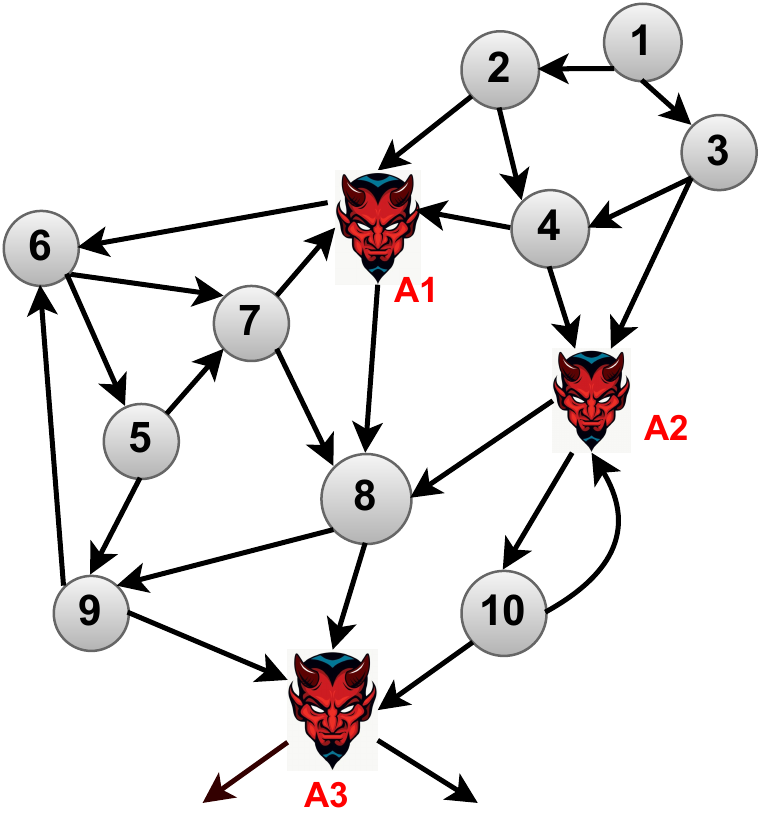}
	\caption{Dandelion++ privacy-subgraph: $A_1$, $A_2$ and $A_3$ represent adversary nodes; rest are benign (honest) nodes.}
	\vspace{-4mm}
	\label{fig:Dan++_adv}
\end{figure}

Similar to Dandelion, Dandelion++ also functions in two phases (stem and fluff) and thus many assumptions remain the same.
However, there are some differences as well; the most significant one with respect to anonymity is the change in structure of the privacy-subgraph in the stem phase.
Dandelion++ constructs an approximate four regular graph as a privacy-subgraph \ie each node ideally should have indegree and outdegree as two. 
In practice, when nodes select two of their neighbors as immediate successors in the privacy-subgraph, it can happen that fewer or more than two nodes select same node as successor. Thus, for each node in the privacy-subgraph, outdegree is guaranteed to be two, but indegree may not always be two. For evaluating anonymity in Dandelion++, we construct an approximate $4$-regular graph and then randomly select a fraction of these nodes to be adversary nodes (ref. Fig. \ref{fig:Dan++_adv}).

Unlike with Dandelion, there are no obvious partitions in Dandelion++. 
This is because in Dandelion, the privacy-subgraph is a perfect line graph (ref. Fig. \ref{fig:Dan_adv}) and thus between any two nodes there exist only one path. In Dandelion++ however, the privacy-subgraph is an approximate four regular graph, thus between any two nodes there normally are multiple paths. An adversary is nevertheless able to reduce the set of possible originators with the following strategies. 

\noindent \textit{\textbf{Combining information from multiple adversary nodes:}}
Since all adversary nodes coordinate among themselves, they can easily identify duplicate transactions (by simply comparing the transaction plaintext). Adversaries only consider the first instance when they received a transaction, ignoring subsequent observations. 
This simple criteria greatly benefits the adversary in deanonymizing transactions. For instance in Fig. \ref{fig:Dan++_adv}, if adversary node $A_3$ receives a transaction it can surely conclude that benign nodes $1$--$4$ cannot be the originators. This is because if any one of them would have been the source, the transaction would be first intercepted by either adversary nodes $A_2$ or $A_1$, and subsequently be ignored by $A_3$. Thus only the remaining benign nodes ($5$--$10$) in the privacy-subgraph are possible originators.

\noindent \textit{\textbf{Incorporating knowledge of the predecessor:}}
We also take into account the immediate predecessor $p$ that forwarded the transaction to the adversary node $j$. We denote as $A_{p-j}$ the event that adversary $j$ received the transaction from predecessor $p$.
The probabilities $P(B_i|A_{p-j})$ can be evaluated (by modifying the eq.\ref{eq:baes_thrm_reduced}) as:

$$
 P(B_i|A_{p-j}) = \frac{P(A_{p-j}|B_i)}{\sum_{k=1}^{N-C} P(A_{p-j}|B_k)} 
$$

\noindent The expression $P(A_{p-j}|B_i)$ is computed as:

$$ 
P(A_{p-j}|B_i) = \sum_{T_P} \frac{1}{2}*\left({\frac{1}{2}*p_f}\right)^{h_{ij}-1}
$$

where $T_P$ is the set of possible paths between benign node $i$ and adversary node $j$ via the predecessor $p$, and $h_{ij}$ is the number of hops between $i$ and $j$ for each valid path.
Note that in Dandelion++ there can be multiple possible paths for a transaction from an originator to reach an adversary node, as the approximate four-regular privacy-subgraph is not a straight line (as was the case with Dandelion).
We thus consider all possible paths (represented as $T_P$) from a benign node $i$ to an adversary node $j$ via predecessor $p$, to calculate the probability of the transaction reaching the adversary node. We sum the probabilities over all valid paths to obtain $P(A_{p-j}|B_i)$. 

As in Dandelion, in Dandelion++ the transaction originator always forwards it to the next hop (\ie the originator never diffuses the transaction). Since each node in the privacy-subgraph has an outdegree two, any one of the two successors is chosen with equal probability of one half. At each subsequent hop in the path the transaction is forwarded with probability $p_f$ (diffused with probability $1-p_f$), but once again any one of the two successors is selected with equal probability. To account for this, for each of the $h_{ij}-1$ intermediate hops in a path from $i$ to $j$ we multiply by a factor $p_f/2$ of choosing each successor in the path.

We now explain with an example how the above strategies reduce the anonymity set. The first strategy \ie sharing information among the adversary nodes is implicitly incorporated in the computation of $P(A_j|B_i)$. For example, in Fig. \ref{fig:Dan++_adv}, $P(A_{A3}|B_1)$, $P(A_{A3}|B_2)$, $P(A_{A3}|B_3)$ and $P(A_{A3} | B_4)$ are \textit{zero} (for all predecessors of $A_3$) because there is no honest path between benign nodes ($1$--$4$) and adversarial node $A_3$ without any other adversary node capturing the transaction first. Thus, $A_3$ can never be the first adversarial node to receive a transaction originated by $1$--$4$. 
Thus, in practice we end up with six possible nodes ($5$--$10$) as originators whenever $A_3$ is the first adversarial node receiving a transaction.  

When taking into account the predecessor,    the possible originators are further reduced, \eg the adversary $A_3$ receives a transaction from node $8$, it knows that $10$ cannot be the originator because there is no path between honest node $10$ and adversary node $A_3$ via the predecessor $8$. This further reduces the anonymity set to five nodes (\ie nodes $5$--$9$) and is incorporated in the analysis, as the value of $P(A_{8-{A_3}}|B_{10})$ is zero.

\subsection{Source Routed}
\label{subsec:LN_approach}
Lightning Network (LN) is a functional and scalable payment channel network with close to ten thousand active nodes. 
This network employs source routing for selecting paths in the network, meaning that the transaction originator decides on the \textit{entire} routing path to reach the node (recipient) with whom it wishes to transact. Each intermediary in the transaction route performs a decryption on the data packet that reveals the next hop in the route, where it forwards the packet.

To evaluate the anonymity offered by LN, we again use our generic approach, which calculates transaction originator probabilities using Bayes theorem.
However, when compared to Dandelion and Dandelion++, modelling of LN has notable differences that need to be accounted for. In hop-by-hop schemes, the routing decisions are taken by each node forwarding the transaction, whereas in source routing the path is determined by the source and dependent on the destination and the path cost (\eg routing fees). Route selection prioritizes paths with lower fees along with considering other path features such as (timelock values). We now describe how we model LN to calculate $P(B_i|A_j)$, describing originator probabilities for an intercepted transaction, which we then use to characterize LN's transaction anonymity. 

\noindent\textit{\textbf{Modelling approach:}}
In LN, whenever a node $i_1$ wishes to transfer some amount to a node $i_2$, the source node $i_1$ computes a ``best'' path along which a payment can be transferred to $i_2$. The path that has the optimal cumulative weight (for the payment transfer) is selected as best path\footnote{The complete LN topology (nodes, edges and weights) is known to each LN node.}.
This weight primarily comprises of the fees charged by individual nodes (at each hop).

From an adversary's perspective, once it is part of a transaction path as intermediary, it tries to establish the originator of the transaction. 
To perform such an analysis, the adversary first computes the best paths for each source-destination pair and then calculates the probability distribution of originators for the intercepted transactions.
To compute the shortest paths it employs Dijkstra's algorithm and the originator probabilities ($P(B_i|A_j)$) are computed using eq. \ref{eq:baes_thrm_reduced}. 
In addition, an adversary can reduce the anonymity set of possible originators by employing the following strategies.





\noindent \textit{\textbf{Incorporating knowledge of the predecessor--successor combinations:}}
Unlike Dandelion++, where we used only the predecessor information, in LN we consider additional information in the form of both predecessor and the successor of the transaction routed through the adversary node. Recall that in LN the source decides the complete transaction path, and thus including the successor further leaks information to the adversary.

For any originator $i$ and adversarial intermediary $j$, we further account for $j$'s predecessor $p$ and successor $s$ in the transaction route. We denote as $A_{p-j-s}$ the event that a transaction received by adversary $j$ from predecessor $p$ is next routed to successor $s$. 
The probabilities $P(B_i|A_{p-j-s})$ are calculated for each subpath $p-j-s$ as:

$$
 P(B_i|A_{p-j-s}) = \frac{P(A_{p-j-s}|B_i)}{\sum_{k=1}^{N-C} P(A_{p-j-s}|B_k)} 
$$

\noindent The expression $P(A_{p-j-s}|B_i)$ is computed as:

$$ P(A_{p-j-s}|B_i) = \frac{SP_{i(p-j-s)}}{SP_i} $$
where $SP_i$ is the total number of shortest paths that originate from $i$ and $SP_{i(p-j-s)}$ is the total number of paths that originate from benign node $i$ and pass through the nodes $p-j-s$, in that order.

We illustrate with a simple example how the inclusion of both the predecessor and successor allows for refining originator probabilities and better identifying the originator.

\begin{figure}[h!]
\vspace{2mm}
	\centering
	\includegraphics[scale=0.7]{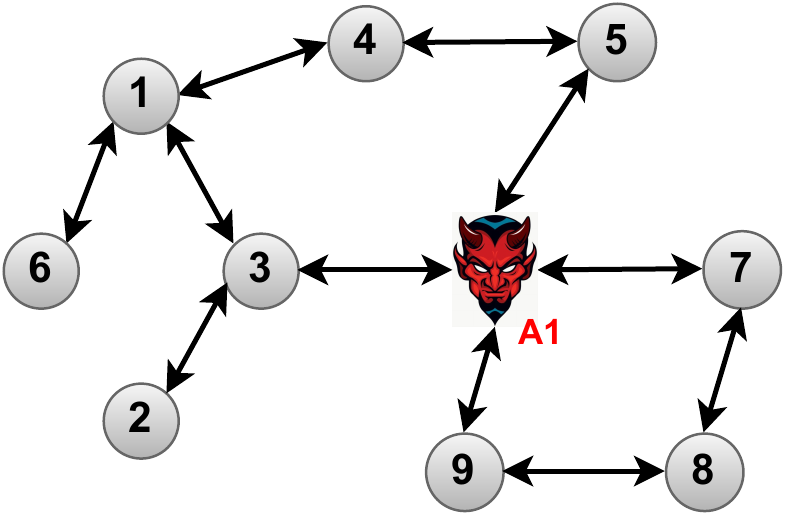}
	\caption{Reducing the anonymity set in LN: Based on the predecessor and the successor of the transaction potential source nodes can be divided into disjoint sets.}
	\vspace{-3mm}
	\label{fig:LN_adv}
\end{figure}

Consider a sample LN topology shown in Fig.~\ref{fig:LN_adv}.
If we take into account the predecessor node from which the transaction has been received, the potential originators are divided into two distinct sets: benign nodes $1$--$6$ (corresponding to predecessors $3$ and $5$) and benign nodes $7$--$9$ (corresponding to predecessors $7$ and $9$). 
Considering predecessor $3$, if we now include the successor $5$ and consider the subpath $3$-$A_1$-$5$, the anonymity set can be further reduced. Assuming that all edge weights are equal, the best paths from nodes $6$, $1$, $4$, and $5$ will not follow the subpath $3$-$A_1$-$5$, but the best paths from nodes $2$ and $3$ will. Thus, if the adversary observes a transaction coming via predecessor $3$ and going to successor $5$, the anonymity set will only contain two possible originator nodes, \ie $2$ and $3$.

Thus, by incorporating the predecessor and successor information of a received transaction, an adversary can reduce the set of potential originators.
Note that in Dandelion and Dandelion++, only the predecessor information is useful to identify the originator, as the successor is unrelated to any originator choice. 

\begin{figure*}
\centering
\begin{subfigure}{.5\textwidth}
  \centering
  \includegraphics[scale=0.5]{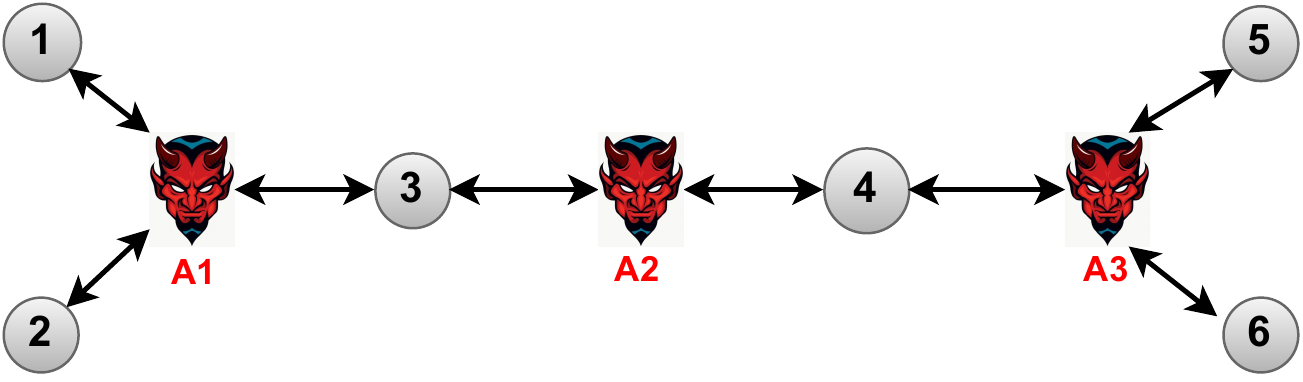}
  \caption{LN path with three adversary nodes $A_1$, $A_2$ and $A_3$ 
  }
  \label{fig:sub1}
\end{subfigure}%
\begin{subfigure}{.5\textwidth}
  \centering
  \includegraphics[scale=0.5]{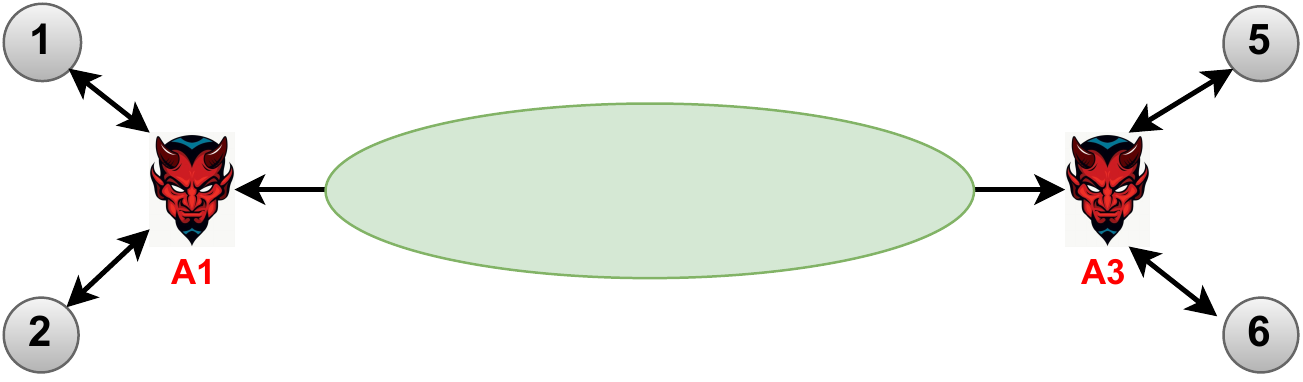}
  \caption{We ignore the all intermediate nodes (including $A_2$) and consider $A_1$, $A_3$ as one adversary node between benign nodes $1,2,5,6$.  }
  \label{fig:sub2}
\end{subfigure}
\vspace{-2mm}
\caption{Subpath computation when multiple adversary nodes are on the same path in LN.}
\label{fig:LN_multiple}
\vspace{-4mm}
\end{figure*}

\noindent \textit{\textbf{Combining information from multiple adversary nodes:}}
\label{subsubsec:Combine_multiple_adv}
We employ an additional simple technique to
combine the information gained by multiple adversaries. We leverage the
the fact that multiple adversary nodes can recognize that they are part of the same path for a particular transaction in the LN. This is possible because, in the LN, every transaction has a unique identifier that all the nodes can see on the said path. Even if such an identifier was not available, adversarial nodes might be able to establish they are on the same path by correlating transaction timing and amount. 

To account for paths containing multiple adversary nodes, we first examine the set of shortest paths and find those with multiple adversaries as intermediaries. 
We then identify the first (closest to the originator) and last (closest to the destination) adversary nodes on paths that contain multiple adversaries.
All the nodes in between these adversary nodes (including other adversary nodes) are ignored. 
Next, we consider the predecessors of the first and successors of the last adversary node and redefine the combined subpaths. For example in Fig.~\ref{fig:LN_multiple}, the combined subpaths for adversary nodes $A_1$ and $A_3$ are 1-($A_1$-$A_3$)-5, 1-($A_1$-$A_3$)-6, 2-($A_1$-$A_3$)-5 and 2-($A_1$-$A_3$)-6.

Combining multiple adversaries may reduce the set of candidate originators compared to individually considering the adversary nodes. 
For example, in Fig.~\ref{fig:LN_multiple}, nodes $3$ and $4$ will not be considered originators as they could not have selected the best path that would involve both nodes $A_1$ and $A_3$. However, if $A_3$ and $A_1$ had individually performed an analysis, they would have considered node $3$ and $4$ as candidate originators. Thus, the combination of $A_1$ and $A_3$ further narrows down the possible originators compared to adversary nodes that would not share their observations.

\noindent \textbf{\textit{Entropy computations for best-$k$ paths:}}
Arguably, LN transactions do not always follow the best path, \textit{e.g.,} when the transaction amount is higher than the payment capacity of the path or when a node in the path is offline, and the transaction has to be rerouted through a different path. Additionally, it is possible to intentionally introduce randomness in the path selection on the client side---rather than selecting by default the best path, clients could select the path by drawing it randomly from a set of good paths.
Introducing randomness in the path selection can lead to enhanced anonymity.

The most accurate way to evaluate the impact of routing randomness due to payment failures is to recreate the exact rerouting behavior specified in the LND path selection algorithm. 
This involves modeling node failure rates as well as keeping a log of past (failed) transactions to compute the bias\_factor (ref. Sec.~\ref{Sec:LN_background}) when retrying after failed payments. In addition, an accurate evaluation would require the real-time directional balances of all channels  \cite{kumble2021comparative}, which are by design not publicly available. This approach is, therefore, non-trivial to implement in practice.

We instead evaluate the impact of randomness in path selection with a simple model that makes abstraction of the various possible sources of randomness and their specific patterns while capturing the uncertainty introduced by the existence of multiple possible paths between a source and a destination. 
The strategy involves computing the $k$-best paths (using Yen's algorithm) between each source-destination pair in LN and considering that any of them may be used for routing the transaction. In these experiments, the originator selects a path uniformly at random among the set of best-$k$ paths instead of deterministically choosing the best path.

Given that rerouted paths are likely in the set of best-$k$ paths (as even after rerouting, the LND routing algorithm majorly optimizes for the same factors as before rerouting), we argue that this strategy approximates rerouting behavior for the purposes of anonymity evaluation. Moreover, if the client waits to first recover the funds from a failed payment before retrying the transaction, the impact of the bias factor is minimal (details in \S\ref{sec:implementation}), making our approximation more realistic. 

Note that the evaluation method for calculating transaction originator probabilities remains the same as in the single best-path scenario, except that in this case, the set of possible paths includes the best-$k$ paths for every source-destination pair.

%% file: sections/Results.tex
\section{Analysis and Results}
\label{sec:setup}

\begin{figure*}
\centering
\begin{subfigure}{.4\textwidth}
  \centering
  \includegraphics[scale=0.45]{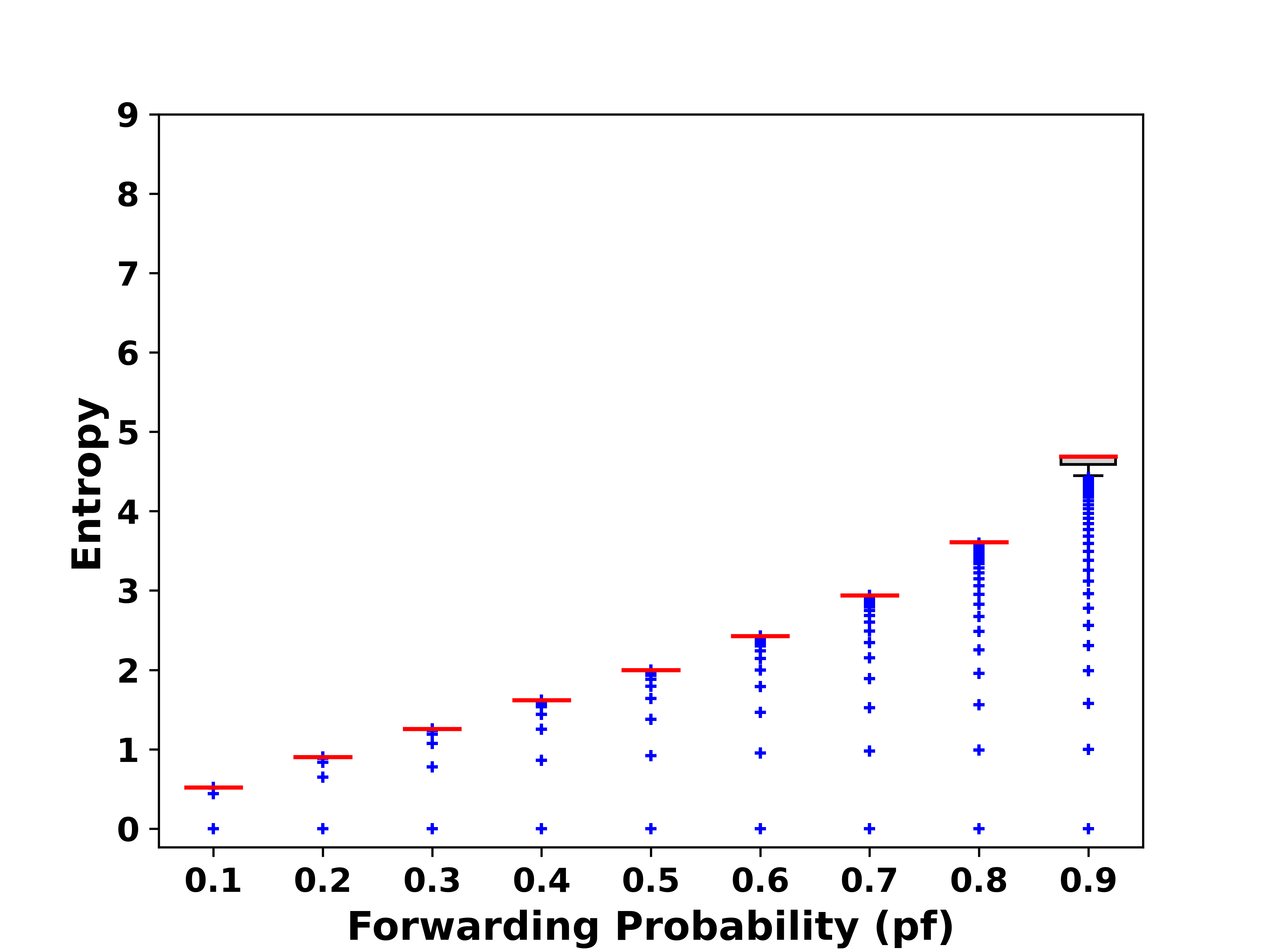}
  \caption{Dandelion}
  \label{fig:Danpf}
\end{subfigure}%
\begin{subfigure}{.4\textwidth}
  \centering
  \includegraphics[scale=0.45]{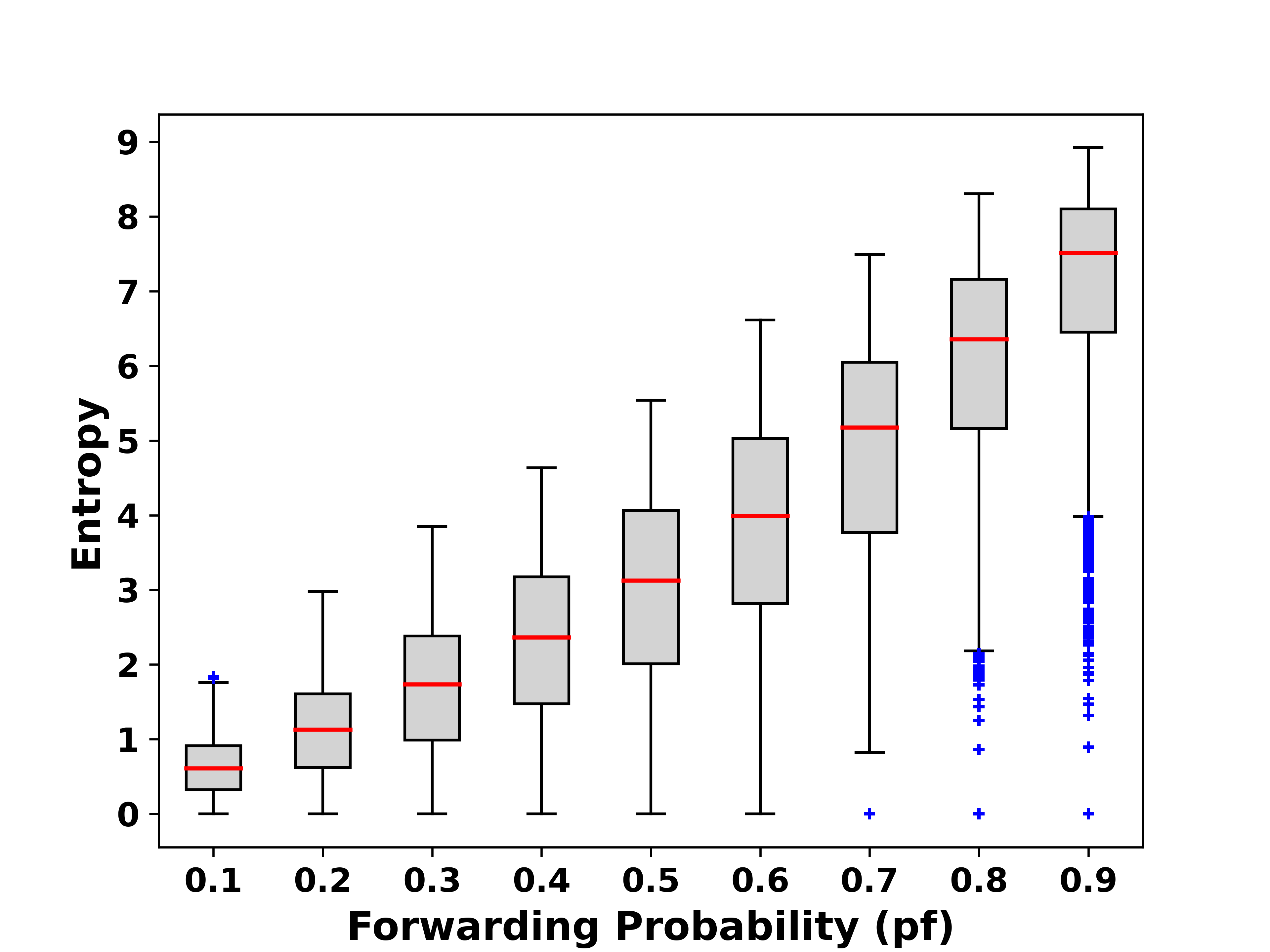}
  \caption{Dandelion++}
  \label{fig:Dan++pf}
\end{subfigure}
\caption{Entropy vs. forwarding probability $p_f$: With increasing the forwarding probability entropy increases.}
\label{fig:pf}
\vspace{-3mm}
\end{figure*}

\begin{figure*}
\centering
\begin{subfigure}{.4\textwidth}
  \centering
  \includegraphics[scale=0.45]{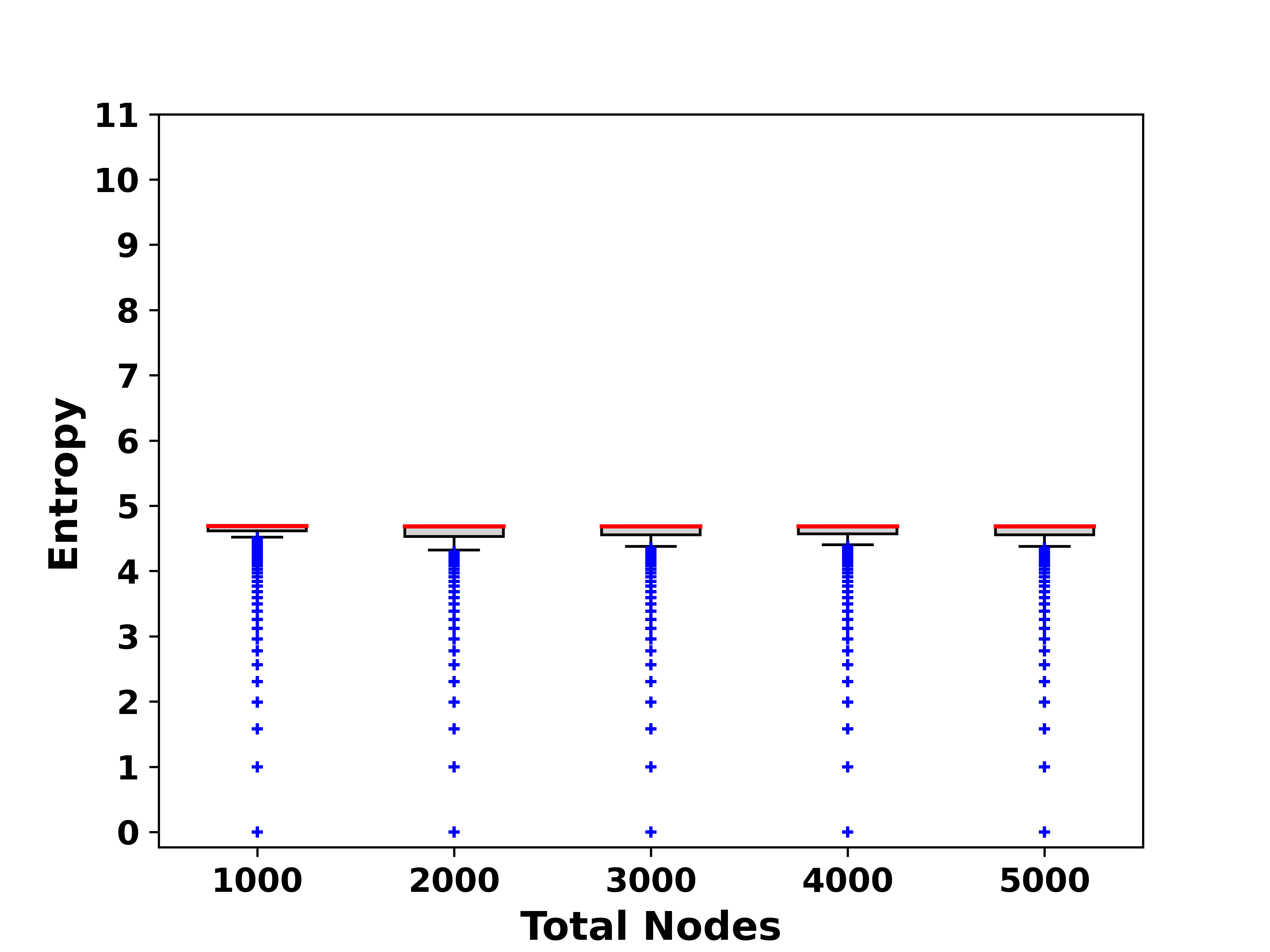}
  \caption{Dandelion}
  \label{fig:DanN}
\end{subfigure}%
\begin{subfigure}{.4\textwidth}
  \centering
  \includegraphics[scale=0.45]{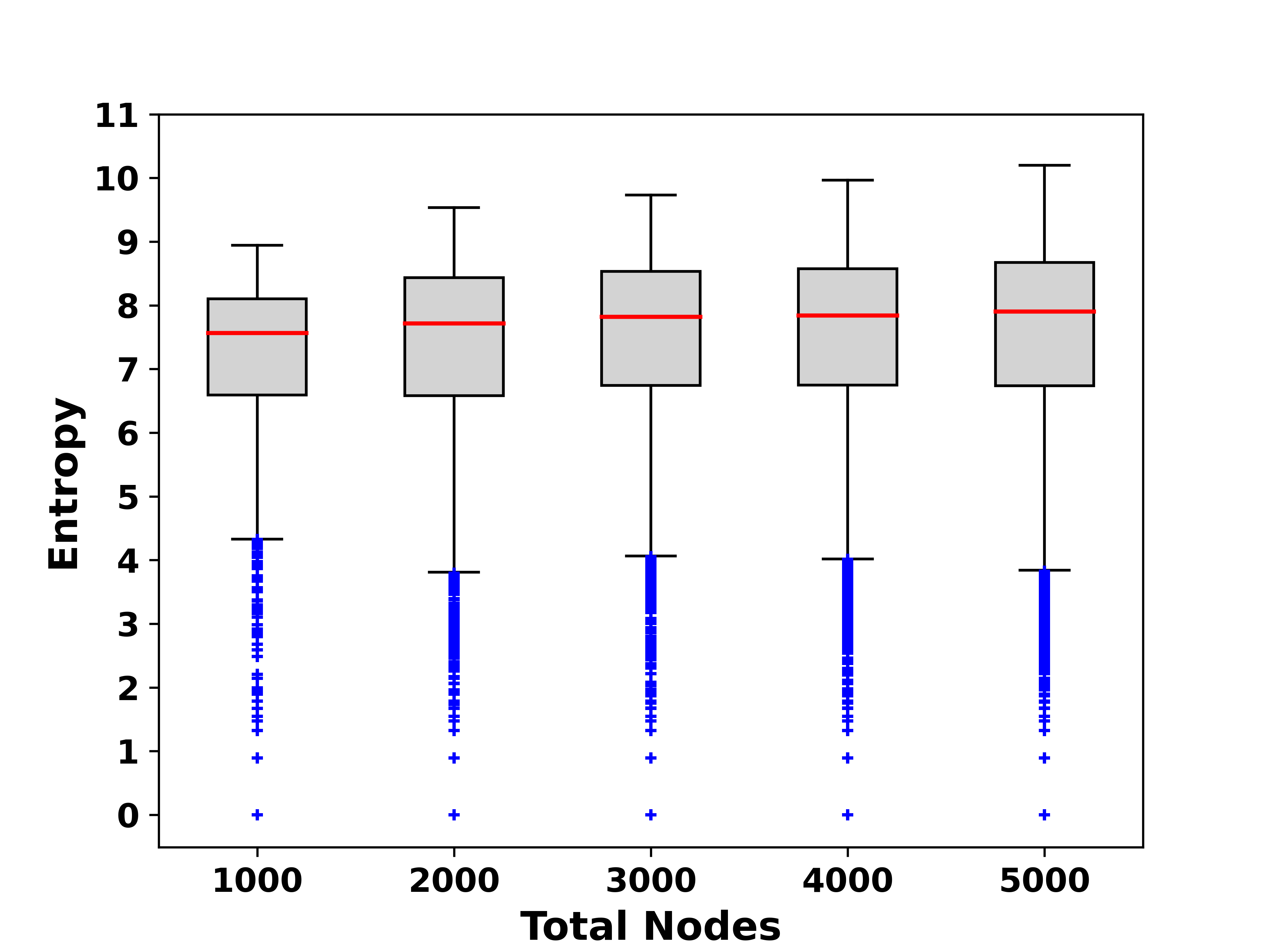}
  \caption{Dandelion++}
  \label{fig:Dan++N}
\end{subfigure}
\vspace{-2mm}
\caption{Entropy vs. total nodes: With increasing the total number of nodes and keeping the adversary nodes as fixed ($1\%$), the entropy does not increase. }
\label{fig:vary_N}
\vspace{-4mm}
\end{figure*}

\begin{figure*}
\centering
\begin{subfigure}{.4\textwidth}
  \centering
  \includegraphics[scale=0.45]{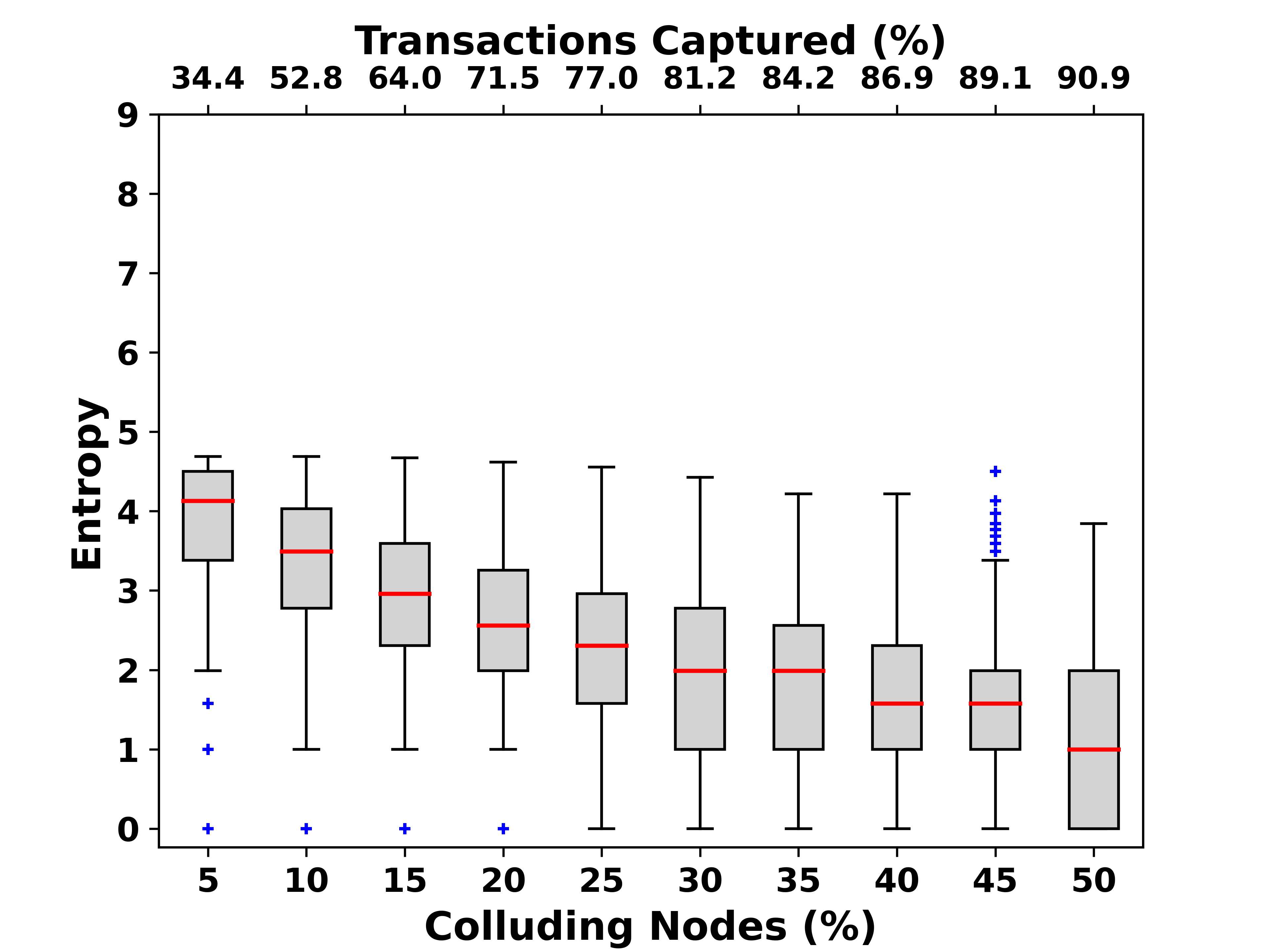}
  \caption{Dandelion}
  \label{fig:Danfrac}
\end{subfigure}%
\begin{subfigure}{.4\textwidth}
  \centering
  \includegraphics[scale=0.45]{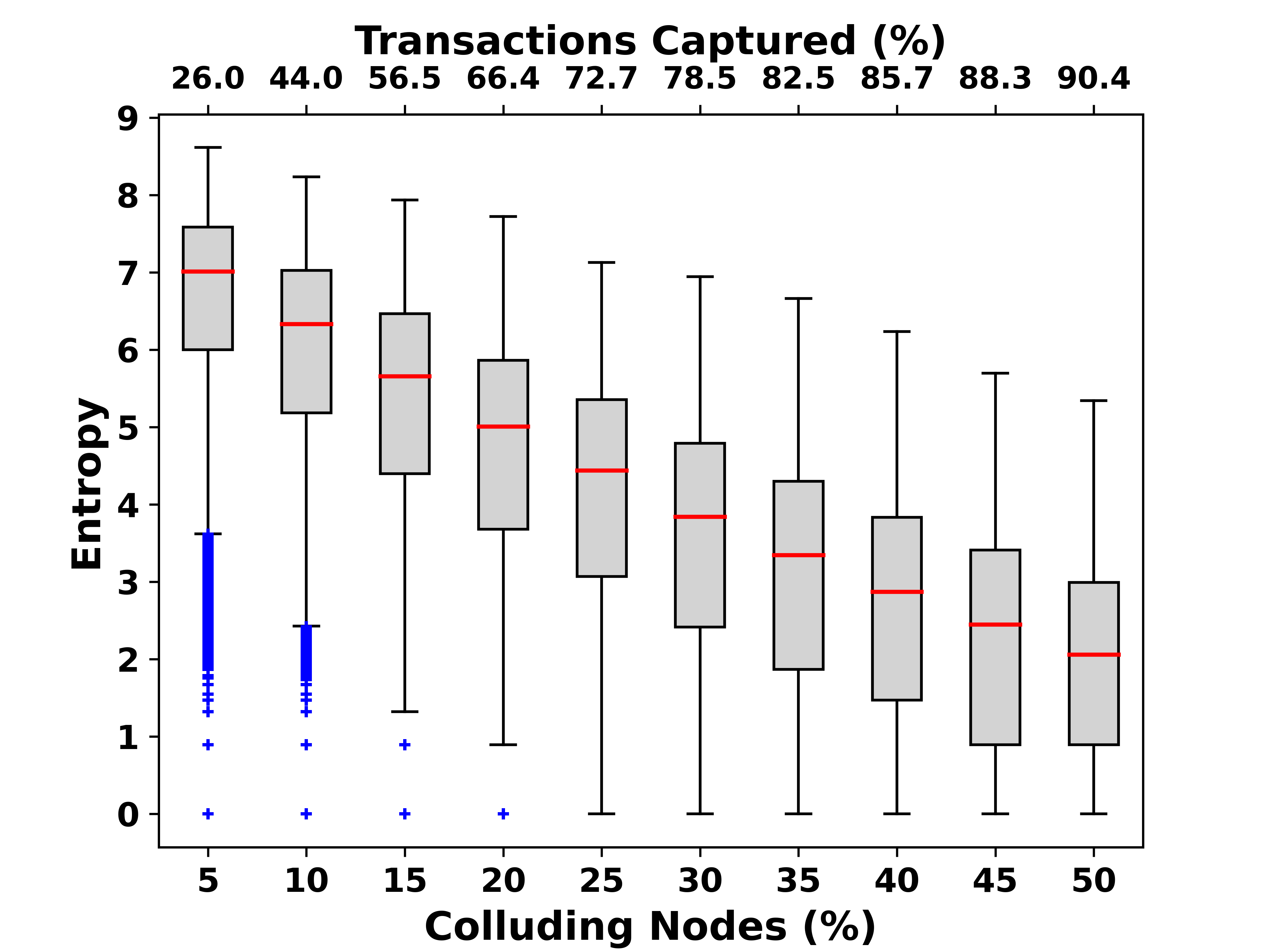}
  \caption{Dandelion++}
  \label{fig:Dan++frac}
\end{subfigure}
\vspace{-2mm}
\caption{Entropy vs. colluding nodes: With increasing the fraction of colluding nodes entropy decreases.}
\label{fig:vary_frac}
\vspace{-4mm}
\end{figure*}

We implemented in Python a simulator to evaluate anonymity for Dandelion, Dandelion++, and LN. At a high level, our simulator generates a network, assigns some nodes to be adversarial, and implements the evaluation approaches presented in the previous sections.
In the simulations, all honest nodes generate an equal
number of transactions. The simulator first calculates the
fraction of transactions captured by the adversary nodes; then, for each intercepted transaction, it computes the entropy of the probability distribution over all possible originators. We represent the results in a box plot that shows the distribution for all the intercepted transactions. These plots indicate the range, median, quartiles, and outliers of the obtained entropy values for the intercepted transactions.

\subsection{Hop-by-hop Routing}
\label{subsec:hop-by-hop-results}

As described in \S \ref{sec:back}, both Dandelion and Dandelion++ route transactions over a privacy subgraph, constructed either as a line graph or as an approximate $4$-regular graph, depending on the scheme.  
The three key variables for the anonymity of such routing schemes are: the forwarding probability ($p_f$); the total number of nodes ($N$), and the number of adversary/colluding nodes ($C$).
We conduct three sets of experiments to isolate the impact of each variable on anonymity:

\begin{enumerate}
    \item Keeping $C$ and $N$ fixed, we varied $p_f$.
    \item Keeping $C$ and $p_f$ fixed, we varied $N$. 
    \item Keeping $N$ and $p_f$ fixed, we increased $C$.
\end{enumerate}

To conduct the experiments, we first generate a random $2$-regular (Dandelion) or $4$-regular (Dandelion++) graph, randomly labeling some nodes as adversaries. Honest nodes then generate transactions, some of which are intercepted by adversaries in the stem phase. For each intercepted transaction, we compute the anonymity of its originator. The previous steps are repeated a thousand times with different adversarial node placements. The box plots contain all the transaction anonymity results computed across these thousand iterations.

\noindent \textbf{\textit{1) Impact of forwarding probability ($p_f$) on anonymity:}}
Previous studies~\cite{dan++} used relatively smaller network topologies (\eg $100$ nodes network) to study the anonymity offered by these schemes. We take the evaluation a step further and simulate larger networks. We constructed privacy-subgraphs with $N=1000$ total nodes considering $C=10$ nodes are corrupted (\ie $1\%$ of $N$, which we re-randomized for each run). We vary $p_f$ from $0.1$ to $0.9$ and for each value of $p_f$ we measured the anonymity for $1000$ simulation runs. 

Fig.~\ref{fig:pf} depicts the results.
We observe that increasing $p_f$ increases anonymity, which is to be expected as a higher $p_f$ increases the transaction's path length, which makes it harder for the adversary to determine the source of the target.
However, we remark that even with very high $p_f=0.9$ and just $1\%$ adversary nodes, the median anonymity value is $5$ bits for Dandelion and $7$ bits for Dandelion++. This means that the effective size of the anonymity set of originators is, respectively, in the order of $32$ and $128$ originators, out of $990$ potential originators. 

\noindent \textbf{\textit{2) Impact of total number of nodes ($N$) on anonymity:}}
In our second set of experiments we fix  $p_f=0.9$ (highest anonymity) and instead increase $N$, while keeping the  fraction of compromised nodes constant at $1\%$, \ie $C=0.01 \cdot N$. For each value of $N$ we again simulate and obtain $1000$ samples, each taken with a random placement of adversaries in the graph. 
A larger $N$ allows a network to increase the anonymity offered to transactions. As shown in Fig. \ref{fig:vary_N}, however, these schemes do not capitalize on the network size increase and instead offer a level of anonymity that remains roughly constant.
Even when $N=5000$, the median (and maximum) entropy value is below $5$ bits for Dandelion \ie an effective set of $32$ originators. 
This is because benign nodes that are far from adversary nodes in the privacy subgraph do not contribute much to the anonymity of transactions intercepted by an adversary. For example, in Dandelion, assuming a benign node $i$ is $35$ hops away from some adversary node $j$, the probability of a transaction generated by $i$ reaching $j$, would be ${p_f}^{34}$, which is very small. Thus, this node would have an insignificant contribution in the final computation of entropy because it is highly unlikely that node $i$ could be the originator of any transaction observed by adversary $j$. Furthermore, if there is another adversarial node in the path between $i$ and $j$, then the adversary can be sure that $i$ is not the originator of a transaction first seen by node $j$. Due to these effects, the increase of $N$ does not translate into an increase of transaction anonymity.   

\noindent \textbf{\textit{3) Impact of adversary nodes ($C$) on anonymity:}}
In the last set of experiments we set $N=1000$ and $p_f=0.9$ and vary $C$ from $5\%$ to $50\%$ of $N$. For each value of $C$ we select $1000$ samples (each with randomly placed adversary nodes).
Our results paint a grim picture---\eg in Dandelion (ref. Fig. \ref{fig:Danfrac}) with $20\%$ adversary nodes, the median entropy is only three bits \ie effective anonymity set of $8$ potential originators (ideally it should have been $850$). Moreover, an adversary with $20\%$ of nodes intercepts $71.5\%$ of all transactions. By comparison Dandelion++ performs better for the same level of compromise. Fig. \ref{fig:Dan++frac} shows that with $20\%$ adversary nodes, the median effective anonymity set is $32$ ($5$ bits), for $>66\%$ of transactions that are intercepted.

\subsection{Source Routing}
\label{subsec:source_routing_results}

To measure the anonymity offered by LN, we required a dataset with its topology since a graph cannot be algorithmically generated to be representative of LN. Thus, we downloaded LN's different real-world
topology snapshots from public sources\cite{lntopology}.
We first present our analysis of LN's December 2018 snapshot, consisting of 1202 nodes and 6196 edges. 
Afterward, we present a longitudinal analysis of how the evolution of LN's topology impacts the anonymity it offers.


Given an LN snapshot, we compute the best paths between every pair of LN nodes in the graph. We then consider that each benign node sends one transaction to every other node in the network. Considering that some nodes are adversarial, we evaluate the anonymity of transactions using the approach presented in \S\ref{subsec:LN_approach} for each of the transactions between benign nodes intercepted by the adversary, \ie transactions whose route includes at least one adversarial intermediary node.

\begin{figure}[h!]
	\centering
    \includegraphics[scale=0.5]{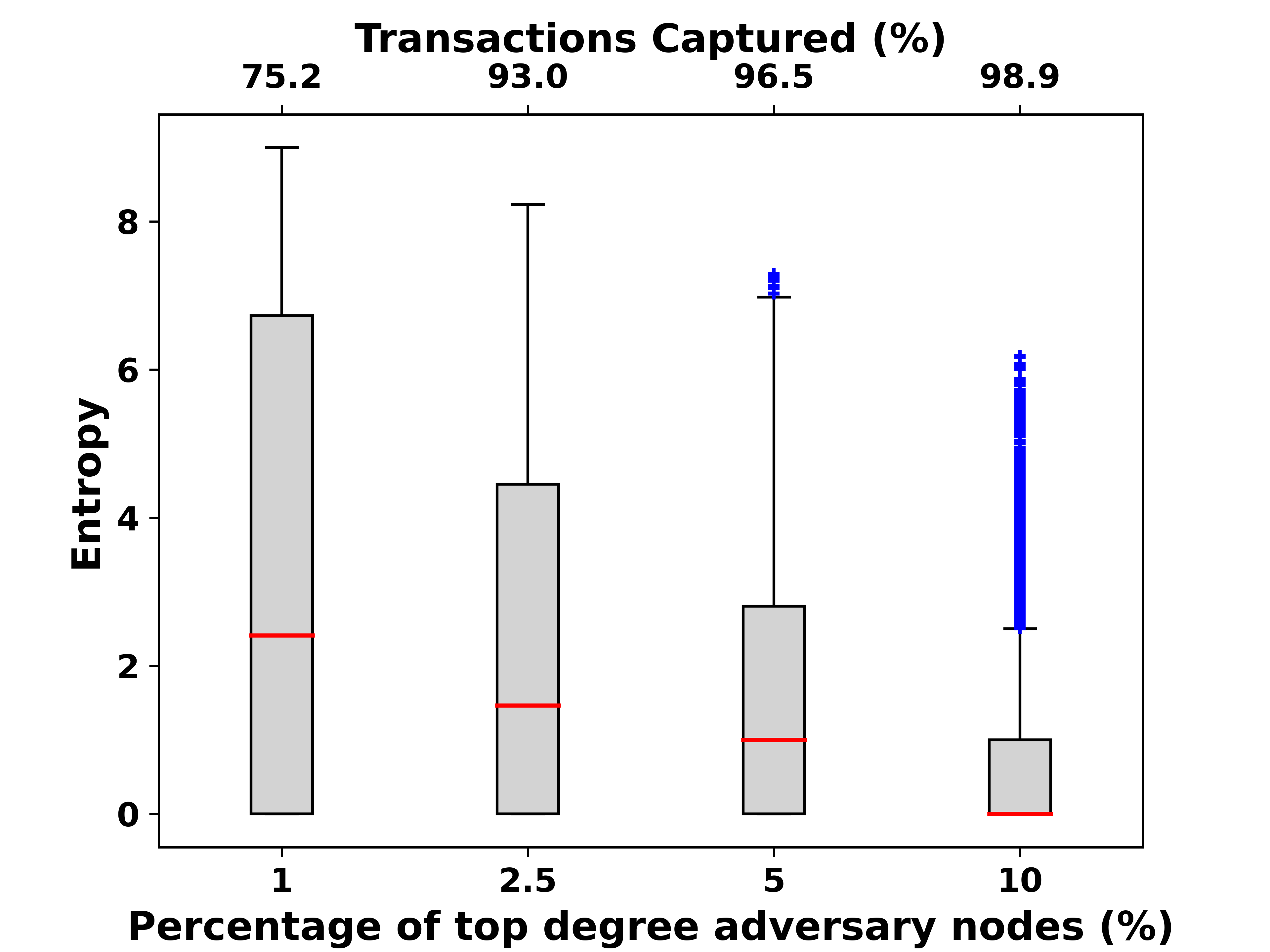}
	\caption{Entropy vs node degree: In each sample, top node degree nodes were selected as adversaries.}
	\label{fig:LN_ND_1200}
	\vspace{-3mm}
\end{figure}

\noindent \textbf{\textit{1) Strategic selection of adversary nodes:}}
A node with many neighbors may observe a large number of transactions that are relayed through it. Controlling such nodes is thus ideal for the adversary. We consider adversaries that control a small number of top-degree nodes to evaluate the threat that large malicious nodes pose to LN users. Our results in Fig.~\ref{fig:LN_ND_1200} show that if the adversary controls the top $1\%$ nodes, the median entropy is $2.5$ bits, and the first quartile is zero. This means that for half of the received transactions, the effective anonymity set is just $6$ possible originators out of twelve hundred LN nodes, and for half of those (a quarter of the total), the transaction originator can be uniquely identified. The adversary can better determine the transaction originator as it controls more nodes, with the median entropy dropping to zero when the top $10\%$ nodes are adversarial.  

High-degree nodes in LN route a disproportionate fraction of transactions, which enables an adversary that controls a few high-degree nodes to intercept significant amounts of transactions.  Controlling $1\%$ of top degree nodes allows the adversary to observe more than $75\%$ of transactions, and this percentage grows to $\approx99\%$ when the adversary controls $10\%$ of top degree nodes. 
We discuss in Sec.~\ref{subsubsec:ln_graph_structure} the implications of some nodes having such a disproportionate influence. 

Additionally, we tested the attack with an adversary that selects nodes with the highest betweenness centrality (rather than a high degree). Nodes with the highest betweenness centrality are those that appear most frequently in the best paths between other nodes, and can thus intercept a large fraction of transactions. The results were very similar to those obtained when selecting adversarial nodes by top degree, indicating that both adversarial strategies are equally effective.  


\noindent \textbf{\textit{2) Measuring anonymity considering best-$k$ paths:}}

\begin{figure}
	\centering
    \includegraphics[scale=0.51]{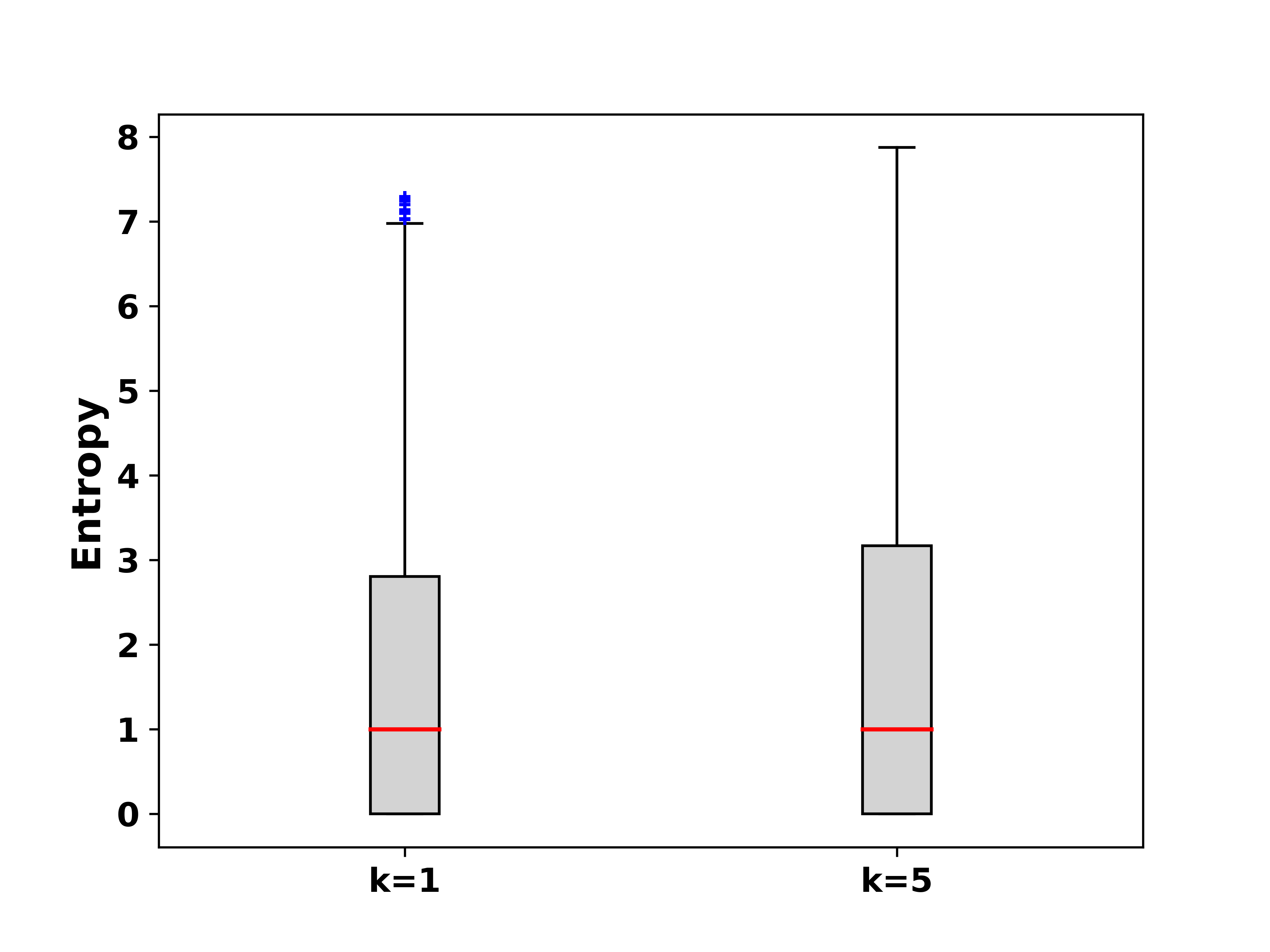}
    \vspace{-2mm}
	\caption{Entropy under best-$k$ paths: Best-$1$ and best-$5$ paths were selected; for both of them, top-$5\%$ degree nodes were selected as adversaries.}
	\label{fig:LN_C_bestk}
	\vspace{-3mm}
\end{figure}

As discussed earlier, a node may not always select the cheapest LN path for routing a transaction, choosing instead randomly from the set of $k$ cheapest paths, some of which may be slightly more expensive than the minimum. This introduces less determinism in LN routing, which should have a positive effect on transaction anonymity. 
To evaluate the impact of this effect, we compare results for networks with identical parameters but different route selection: the first network has deterministic path selection ($k=1$), where the best path is always chosen, and a second network that selects the path uniformly at random among the best $5$ paths (computed using Yen's algorithm).
We observe that increasing the possible paths does not have much of an impact on anonymity. For instance, Fig.~\ref{fig:LN_C_bestk} compares the two scenarios considering the top $5\%$ degree nodes as adversaries. In both cases, the median entropy is just $1$ bit. We saw a similar trend for other adversarial fractions as well.

\noindent \textbf{\textit{3) Longitudinal analysis:}}
The LN has grown significantly in the last few years. Compared to the December 2018 snapshot used in the previous experiments, by May 2021 the LN includes more than $9300$ nodes and $\approx52K$ edges, with the largest connected component containing more than $8100$ nodes and $\approx51K$ edges. Here we examine if the scaling of LN has had a positive impact (as one might expect) on the anonymity offered to transactions. 
To that end, we perform a longitudinal analysis for different LN snapshots (for the years $2018$ to $2021$). As previously mentioned, the initial LN topology consisted of $1202$ nodes and $6196$ edges. But, in $2019$, nodes increased by $2724$ and edges by $37490$; in $2020$, the network further grew and had a total of $5254$ nodes and $60970$ edges.

\begin{figure}[h!]
	\centering
    \includegraphics[scale=0.5]{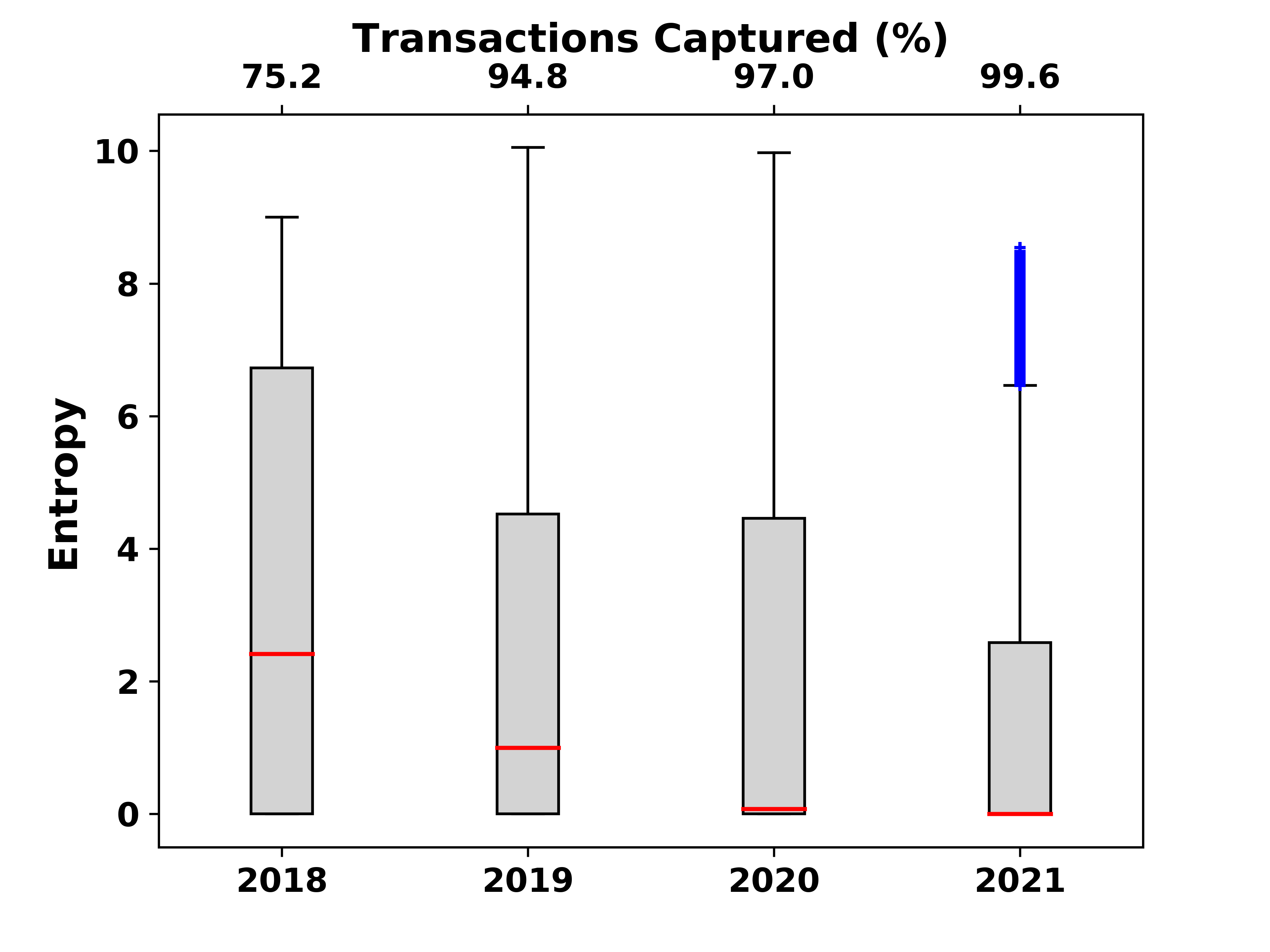}
	\caption{Entropy variation over time (in years): We select top $1\%$ degree nodes as adversaries and compute the entropy.}
	\label{fig:LN_long}
	\vspace{-3mm}
\end{figure}

Once again, we use node degree as criteria to select the adversary nodes, choosing $1\%$ of the top degree nodes as adversarial for all the snapshots and performing the analysis. 
Contrary to expectation, our results reveal that transaction anonymity decreases with the growth of the network, whereas the fraction of transactions intercepted by the adversary nodes increases (ref. Fig. \ref{fig:LN_long}). \textit{For the $2021$ snapshot, an adversary controlling the top $1\%$  nodes can intercept more than $99\%$ of the total transactions and  completely deanonymize half of them (median entropy is $0$ bits).} Even for most of the remaining transactions the entropy is low (less than $3$ bits), barring a few lucky ones. This means that a few strategically chosen adversary nodes can deanonymize a major fraction of all the transactions in the network.
Such observation can be attributed to the fact that LN has a scale-free topology \ie a few influential nodes capture a large fraction of routing paths. 
For instance, when we analyzed the $2021$ topology, we found 6026 nodes with zero centrality, 1483 nodes with
centrality between zero and one and only 604 nodes with centrality values greater than one, highlighting the disproportionate  routing centrality (see \ref{subsubsec:ln_graph_structure} for more details).

\noindent \textbf{\textit{4) Impact of transaction amount on anonymity:}}

Throughout our analysis, we assume that clients generate transactions of a minimal amount. Thus for computing the shortest paths between any two given nodes in LN, we considered all possible paths (consisting of all channels with non-zero capacity) in the LN snapshot. 
Thus our results provide an upper bound on the paths that can be selected for Dijkstra's computation. 

However, in practice, the adversary can perform an even more precise analysis if it considers the \textit{exact amount} being transferred in the transaction.
Transactions with larger amounts can only be relayed via a subset of all possible paths \ie only those paths that have payment capacity at least equal to the transaction's amount. This reduces the total number of paths to be analyzed and, thus, the anonymity set.
Moreover, it might also happen that for such bigger transactions, from some nodes there are no paths with adequate capacity to perform the transaction. This could further reduce the anonymity set for higher amount transactions.

Thus, we conducted experiments considering the latest LN topology and different transaction amounts and evaluated their impact on transaction anonymity. In LN, the total channel capacity (between the two peers) is publicly known, but the exact capacity in each direction is unknown. Thus in our analysis, we assume that half of the total capacity of the channel is available in either direction. Using this assumption, we filter out channels whose capacity does not support a given transaction amount and perform the analysis on the remaining LN graph.
This leads to network partitioning---one big largest connected component (LCC) and then multiple small components (with a maximum of three nodes). For instance, our analysis of $\$10$ transactions amount yields the LCC of $7211$ nodes and other $1985$ small components. As the transaction amount increases, the size of the LCC decreases: for $\$50$--$4356$ nodes, $\$100$--$3196$ nodes, and $\$500$--$1468$ nodes.  

\begin{figure}[h!]
	\centering
    \includegraphics[scale=0.5]{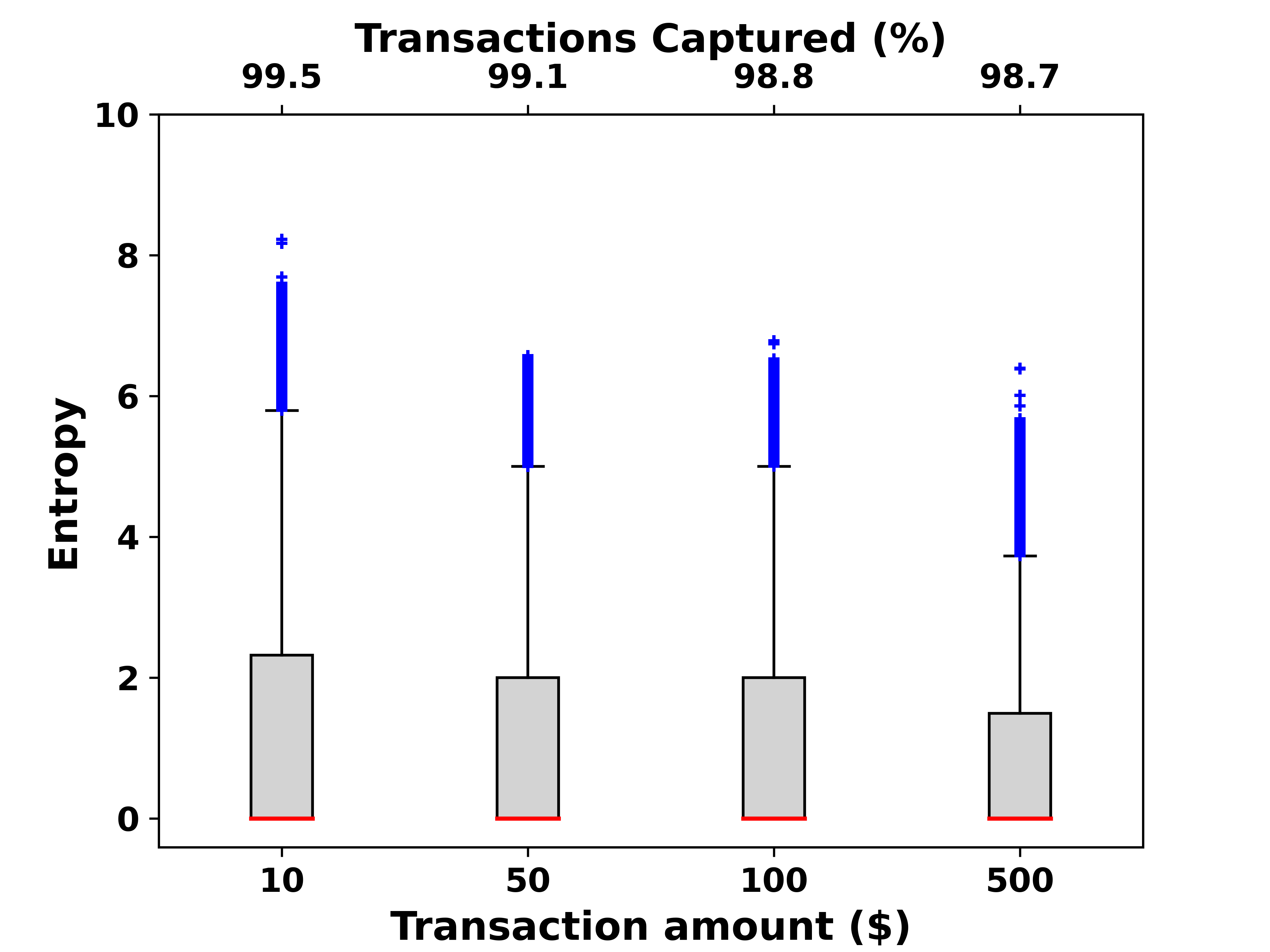}
	\caption{Entropy vs transaction amount: We select top $1\%$ degree nodes as adversaries and compute the entropy.}
	\label{fig:LN_cost}
	\vspace{-3mm}
\end{figure}

Fig. \ref{fig:LN_cost} depicts the entropy distribution for the various transaction amounts when the top $1\%$ high degree nodes are selected as the adversary in the latest LN topology. The median entropy is zero for all amounts, and as expected, increasing transaction amount reduces its anonymity due to the reduction in possible transaction routes. 


\section{Implementation Details}
\label{sec:implementation}

\color{black}
We now present details about our simulator and how we created topologies and implemented routing for Dandelion, Dandelion++, and LN.

\noindent \textbf{Simulator:} We developed our own simulator to evaluate the three considered schemes under different settings.
Our simulator consists of the following components: (1) Pre-processing scripts for generating (or importing) topologies. 
(2) Configuration files that contain information about topology and tunable parameters (\eg $p_f$, $N$, $C$, sample size \etc). (3) The compute engine that instantiates the network and computes the intermediate results that are logged to files. (4) Scripts to process the logs, calculate results (\eg entropy) and plot the graphs.
The simulator is written in Python and has about $3$K lines of code. The source code of our simulator is public and can be found at \cite{Simulator-code}.

In general, our analysis required heavy computations on graph topologies \eg in the LN topology consisting of more than $9$K nodes and more than $51$K edges; we needed to compute Dijkstra shortest paths between all pairs of nodes.  
Thus, we ensured that our compute engine is a highly paralleled program that spawns different threads for computing originator probabilities and, eventually, the entropy. 
To achieve parallelization, we used \texttt{multiprocessing} Python library.  We performed our analysis on two servers with Intel Xeon processors, each with $128$ GB RAM and $20$ physical cores. 

\noindent \textbf{Topology:} The topologies for Dandelion and Dandelion++ are constructed with their specified constraints \ie line graph for Dandelion and quasi $4$-regular graph for Dandelion++, using the \texttt{networkx} Python library. For LN, we used the real-world snapshots that were obtained from ~\cite{lntopology}. We extract the information about channels, nodes, and channel policies (fees, timelock values, \emph{etc.}) from these snapshots for our analysis.

\noindent \textbf{Routing:} The routing policy for Dandelion(++) is implemented as specified in the original paper, with each hop tossing a biased coin to decide whether to forward the transaction to the next hop in the privacy subgraph or broadcast (diffuse) it to the network. 

LN, however, has multiple client implementations (LND, c-lightning, and eclair) with variations in how they construct paths~\cite{tochner2019hijacking,kumble2021comparative}. For our analysis, we used LND's implementation, as the majority of the nodes in the LN ($>90\%$) use it~\cite{rohrer2020counting,mizrahi2021congestion}. The routing algorithms of the other implementations can be easily incorporated by modifying the weight function before calculating the best paths using Dijkstra. 
Note that the results of a related LN  evaluation~\cite{kumble2021lightning} reveal that anonymity sets are almost identical with and without considering other less used client implementations, \textit{i.e.,} eclair and c-lightning (ref. Sec.5.4 in ~\cite{kumble2021lightning}). Accounting for these other client implementations should thus have no impact on the overall conclusions on LN's anonymity.

More specifically, in the implementation, we use the LND cost function when selecting transaction paths. 
In almost all our analyses (except for the one that studies different transaction amounts), we assume the transaction value to be of the minimum amount so as to consider the maximum number of possible paths for the transaction. Note that this constitutes the worst-case scenario for an adversary trying to identify the transaction originator, who has an easier task identifying the originators of transactions with higher amounts for which some paths can be eliminated due to having insufficient capacity. 

One simplification we make is that we do not implement LND's rerouting algorithm for failed transactions,\footnote{Rerouting is generally challenging to model accurately in practice. 
It requires obtaining historical payment failure data (for each node) and the real-time channel balances to anticipate payment failures~\cite{kumble2021comparative}, which are publicly not known.} partly because considering the minimum amount for transactions minimizes the chances of payment failures.  
In cases where payment does fail, a client has to wait for a minimum duration of 40 bitcoin blocks (going as high as 144
blocks) \textit{i.e.,} about a minimum of 6 hours to get back the invested funds in HTLCs.\footnote{It takes roughly 10 minutes to mine one bitcoin block.}
If the client waits to get the funds back from the network before attempting to conduct the transaction again, the rerouted path is likely to be the same as the original path. 
This is because, after six hours, the value of the bias for the channel that failed the transaction becomes almost equal to the bias without a payment failure (98.61\% of the original value after 6 hours); meaning that the impact of payment failures on subsequent path selection is rather minimal.\footnote{If the client wishes not to wait, he would ideally prefer using some alternative means to pay.}

%% file: sections/Discussion.tex
\section{Discussion}

\subsection{Graph Learning Attacks}
\label{subsec:grpah_learn}

In order to successfully deanonymize the source of a transaction, the adversary must know the underlying network topology: the complete network graph in case of source routing  \ie LN; and the privacy-subgraph, \ie line graph for Dandelion and $4$-regular graph for Dandelion++, in hop-by-hop routing schemes.
Obtaining LN topology is trivial, as every node maintains a completely up-to-date topology. A locally available topology is essential for computing paths and routing payments through the network.
Obtaining a privacy subgraph is relatively harder, but still possible.
In Dandelion, $x\%$ adversary nodes can easily infer the positions of about $2x\%$ benign nodes, as they know their immediate successor and predecessor in the line graph. Similarly, in Dandelion++ the information of approximately $4x\%$ nodes is directly available.
The adversary can employ different techniques to infer the rest of the privacy subgraph. One such approach consists of sending transactions to honest nodes whose successors in the privacy subgraph are yet to be identified, then observing who diffuses those transactions~\cite{fanti2017deanonymization}. 
The transactions sent to honest nodes to learn the graph may be generated by the adversary, but not necessarily. The adversary may simply relay any transactions generated by other nodes that are sent to its node.

When many transactions are routed via an honest node, the distribution of nodes that diffuse these transactions allows inferring the successors of that target node in the privacy subgraph.
This is because the successors' frequency of diffusion is directly dependent on the number of hops in the privacy subgraph between the target and the successor, being highest for immediate successors and decaying geometrically with the number of hops. The adversary can combine this information with knowledge of the bitcoin graph\footnote{Although the bitcoin graph is not publicly available. There are various researches~\cite{delgado2019txprobe,miller2015discovering,franzoni2021atom,grundmann2018exploiting,eisenbarth2021comprehensive} that demonstrate how the bitcoin graph can be reconstructed with high accuracy.} to significantly narrow down the possible graph neighbours and strengthen its inference, taking into account that only the $8$ immediate neighbors of a node in the bitcoin graph are potential successors in the privacy subgraph (since privacy subgraph is derived from the bitcoin graph). 

We simulated this approach and observed that typically $2$ of the $8$ bitcoin graph neighbours of a node diffuse the most transactions, and can thus be easily identified as the successors of the honest node in the privacy subgraph. We were able to learn more than $98.5\%$ of the privacy subgraph of Dandelion++ when we sent $100$ transactions per honest node. In Appendix~\ref{app:priv-learn}, we explain our approach in a step-by-step manner and also present some additional techniques that further optimize the overall approach, lowering the number of transactions required to learn the graph. Overall, by analyzing a sufficient number of transactions, the adversary can reconstruct the privacy subgraph.

\color{black}
In practice, the adversary may have uncertainty about the placement of some nodes in the graph, \textit{e.g.,} due to churn in the network the position of new nodes may not be known. A simple strategy that can be adopted by the adversary is to consider all such nodes as potential originators for every transaction. If the fraction of such nodes is low, the overall entropy will not vary significantly.
For instance, when the complete graph is perfectly known and $10\%$ of the nodes are adversarial, we obtain the average entropy of 7 bits (ref. \S\ref{subsec:hop-by-hop-results}). If the adversary is uncertain about the placement of $2\%$ of the nodes, the observed entropy increases slightly to 7.15 bits.
Note that if a significant fraction of the graph is unknown, then the anonymity will significantly increase (due to the lack of information on the majority of nodes). A scenario with a large unknown fraction is however less likely, given the effectiveness of the proposed approach to infer the privacy subgraph with a limited number of transactions.

\subsection{Impact of Churn}
\label{subsubsec:churn}

Churn refers to nodes joining and leaving the network, which are events that happen on a routine basis in distributed peer-to-peer networks. We thus discuss here the impact of churn on our methods. 
In our analysis, we consider a passive adversary that controls a fraction of nodes used to record observed transactions. The adversary attempts to identify transaction originators by combining malicious node observations with information about the network graph used for routing transactions. The current network graph is either downloaded, as in the case of LN, or inferred, as in the case of Dandelion(++). 

With respect to churn, two scenarios are relevant: (1) a new node joins, and (2) an existing node leaves the network. If a new node joins the network \textit{after} the adversary has observed a transaction, the new node should not be considered a potential originator of that transaction, and this is also what we do. When an existing node leaves, it should ideally be eliminated from the set of potential originators for all transactions generated after the node went offline. However, our current analysis still considers it as the potential originator. Note that assuming all nodes are online is a ``worst-case'' scenario from the adversary's perspective, as it considers more nodes as potential originators than the ones that are currently live and available, increasing the anonymity set. Even then, we observe low entropy values for most of the transactions. Given a reliable way to obtain liveness information for all nodes, the analysis can be further refined with such information, leading to even lower entropy. This would not change but rather strengthen the overall conclusions of our work that the studied schemes provide low anonymity. 
Obtaining the liveness information for LN is relatively easy as the topology information is publicly available and regularly updated. On the other hand, obtaining such liveness information for the bitcoin P2P graph in real-time is far from trivial as the topology information is not published and is instead learned over time in a distributed manner.

\subsection{Lightning Network Graph Structure}
\label{subsubsec:ln_graph_structure}
We observed that the vast majority of nodes in the LN (for different snapshots) have (betweenness) centrality zero or less than one, while relatively few nodes have overwhelmingly large values of centrality. This indicates that there is ``routing centralization'' within LN \ie a few nodes capture a large fraction of network paths, a fact already noted in prior studies \cite{rohrer2019discharged,tochner2019hijacking}. The analysis by Rohrer et al.~\cite{rohrer2019discharged} of an actual LN topology snapshot of 2019 highlighted that the node degree distribution of LN follows power-law, suggesting a \textit{scale-free} network structure. They also demonstrated that LN can also be classified as \textit{small-world} network.
\ie nodes tend to cluster and have a high density of edges in their cluster.
Our results indicate that the growth of LN ($>9$K nodes presently) has not diminished ``routing centralization'' but rather exacerbated it.
We thus turn to ask whether a change in graph structure would improve the anonymity offered by LN, \ie, if the LN graph was more balanced, with roughly similar degrees for all nodes instead of node degrees following a power law, would the network provide better anonymity? 

\noindent \textbf{\textit{Impact of change in graph structure on LN:}}
We create random graphs with the constraint that the average degree ($k$) of nodes should remain constant (\eg $k$=$5$). We assign weights to the edges following a distribution similar to the actual LN (the average fee associated with a channel is $1000$ millisatoshi) and  compute the shortest paths between all pairs of nodes. We select the top $1\%$ centrality nodes as adversary nodes and compute anonymity for the transactions intercepted by the adversary. 

\begin{figure}[h!]
	\centering
	\includegraphics[scale=0.5]{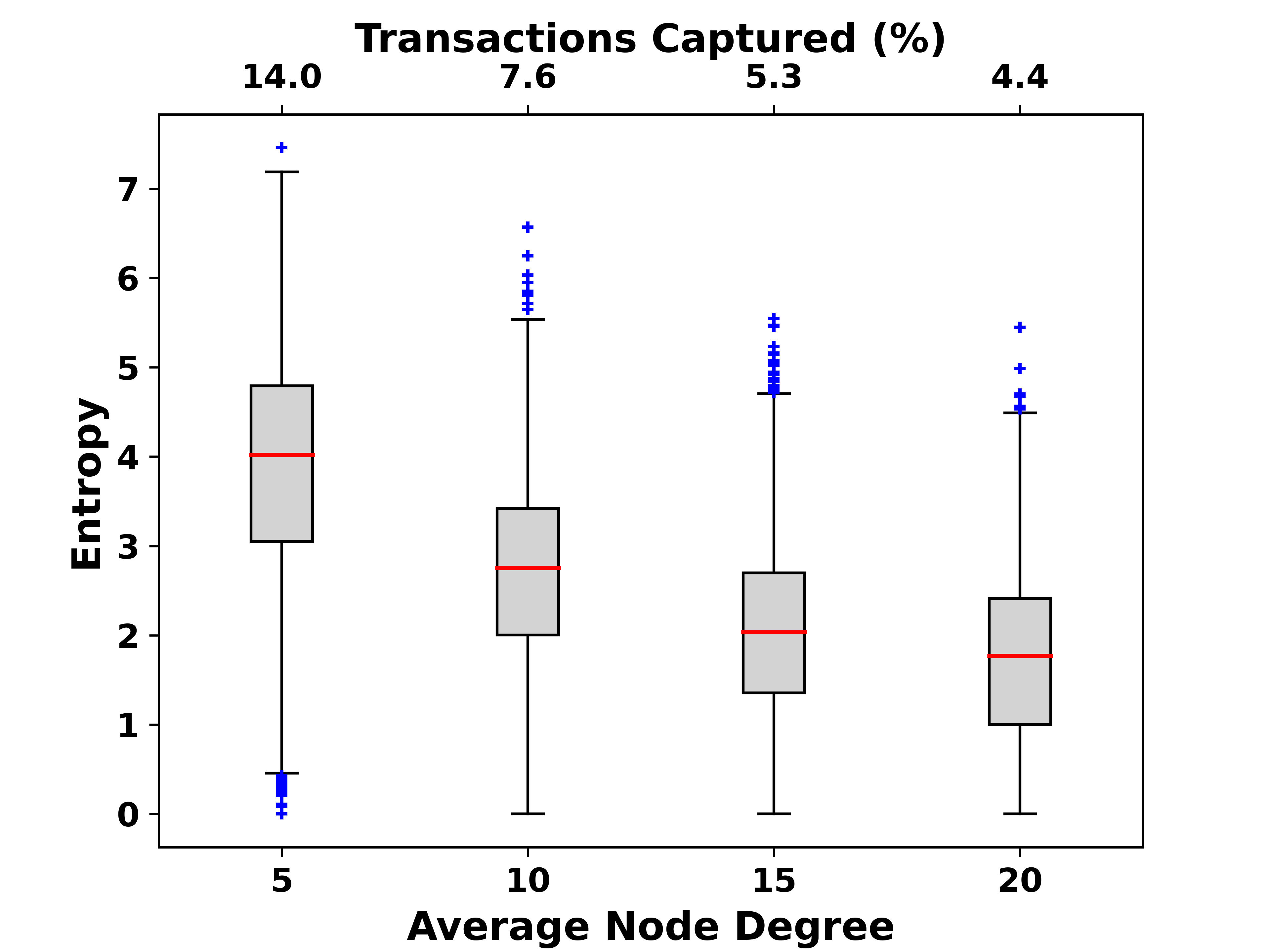}
	\caption{Entropy for random $k$-degree graph: For $1000$ node topology, we increased the average node degree, kept the edge weights roughly the same, and study its impact on entropy (with top $1\%$ centrality nodes as an adversary).
	}
	\label{fig:LN_random_1000}
	\vspace{-3mm}
\end{figure}

We observe that transaction anonymity is somewhat higher in these balanced graphs (\eg median $5$ bits of entropy with high standard deviation for $k$=$5$) in comparison to the actual LN topology (median zero with low standard deviation in the latest snapshot). 
The overall entropy values are however still low and allow the adversary to significantly narrow down the identity of transaction originators.
On the other hand, despite the overall entropy being low, in random networks, the adversary intercepts a smaller fraction of transactions compared to the LN real topology snapshots (ref. Fig. \ref{fig:LN_random_1000}). This is because in random graphs all the nodes have similar node degrees and their centrality values are balanced (almost none has zero centrality). Thus, a selection of top centrality nodes does not provide much advantage compared to a selection of random nodes -- and in both cases the number of intercepted transactions is moderate.
This is unlike the actual LN snapshots, where $\approx64\%$ nodes have zero centrality---which allows those high-centrality nodes to become intermediaries for a disproportionate fraction of transactions. 

\subsection{Comparing Results with Dandelion(++)} 

In our analysis, we use a different metric than Dandelion(++) to measure the offered anonymity. 
We use entropy, a metric that accounts for all possible guesses (\emph{i.e.,} all non-adversarial nodes); Dandelion(++), on the other hand, uses precision-recall metrics that only focuses on the best possible guess (\emph{e.g.,} predecessor is the originator), ignoring any specific routing information known to the adversary. Thus a direct comparison between our analysis and the Dandelion(++) is not straightforward.


Moreover, the analysis in Dandelion is not specific to an exact deanonymization strategy deployed by the adversary. Instead, the results present bounds on the precision and recall an adversary would obtain when it follows any deanonymization strategy. These strategies are defined as a mapping function that always attempts to assign any given transaction to a single originator. Such assumptions do not consider the cases where even though the adversary cannot correctly guess the exact sender with sufficient confidence but can know with high probability that the sender is among a small set of senders. For instance, let us consider that a node has a probability of being the originator as 0.5 and another one has 0.49, with all the remaining nodes having a combined probability of 0.01. In such a case, the Dandelion(++) matching function of considering the highest probability node would yield the node with 0.5 probability as the originator, whereas a more accurate assessment would have been that the sender is one of the two nodes with a very high probability. Such assessments are provided by entropy as it considers the probability of all nodes being the potential originator and then reports the effective size of the anonymity set. 

Thus, overall our results do not directly contest the analysis presented by Dandelion(++) as they provide bounds rather than calculating transaction anonymity sets.
The bounds provided by Dandelion(++) may still be applicable to the matching strategies (\emph{e.g.,} `predecessor is the originator'). However, to measure the anonymity of individual transactions, their metric (and analysis) is non-trivial to contextualize. To that end, entropy reports the anonymity sets per transaction, which is simpler to comprehend and clearly demonstrates that overall anonymity provided to transactions is low.

Interestingly, some of our inferences are in line with what was reported in Dandelion++. We observe that increasing the node degree in the privacy subgraph would lead to larger entropy and thus anonymity. As discussed next, similar findings were reported by Dandelion++, where the authors mention that if the privacy subgraph is known, then increasing the node degree in the privacy subgraph leads to a lower precision for the adversary and thus results in increased anonymity. 




\subsection{Improving Dandelion's Anonymity}
\label{subsec:Improving_Dan}
Our experimental results demonstrate that Dandelion and Dandelion++ do not offer large anonymity sets against colluding nodes. An important reason for this is that these schemes use restricted routing. In Dandelion, a node can forward to just one neighbor, whereas in Dandelion++, it can forward to just two neighbors.
Our analysis demonstrates that changing the privacy subgraph from line to $4$-regular results in better overall entropy. 
Thus, we investigate if increasing the node degree in the privacy subgraph leads to further anonymity improvements.
Since the privacy subgraph is derived from the bitcoin graph, which is itself $16$-regular, this is the maximum degree that any node in the privacy subgraph can have. Thus, we constructed a $16$-regular graph of $1000$ nodes and computed the entropy as described in Sec.~\ref{subsec:LN_approach}. We observed that with $10\%$ adversary nodes, the median entropy value was $7$ bit (equivalent to $128$ possible senders) in the $16$-regular privacy graph and $4.5$ bit ($23$ possible senders) in the 4-regular graph.\footnote{To bound the computations, we considered only $5$-hop paths from the source to the adversary nodes.}

Thus, to achieve better anonymity, hop-by-hop schemes like Dandelion should use the $16$-regular bitcoin graph \textit{as-is} for the privacy subgraph instead of restricting it to a line or 4-regular graph.
Notably, in this case, the adversary already knows the privacy subgraph without the help of any additional mechanisms (as described in Sec.~\ref{subsec:grpah_learn}), but is still less able to effectively deanonymize transaction originators. 

\color{black}
We note that the authors of Dandelion also considered forwarding the transaction to multiple bitcoin nodes (\textit{i.e.,} more than two nodes). 
However, they assume that the privacy subgraph will not be known to the adversary. With that assumption, they argue that ``diffusion by proxy'' (forwarding to any bitcoin node) is not the best strategy. It intuitively provides too many paths for the transactions to reach the adversary, and thus the relative mixing of packets is very less. Hence they suggest using a line (instead of a complete) graph as the privacy subgraph. Line graphs limit exposure to adversary nodes, and at the same time mixing would be high as more transactions would follow the same path.

However, in Dandelion++, the authors relax the assumption of not knowing the privacy subgraph and analyze the two cases---when the privacy subgraph is unknown and when it is known. They conclude that if the privacy subgraph is unknown to the adversary, Dandelion's analysis still holds. But, with a known privacy subgraph, the trend reverses. As shown in our analysis in this paper, the adversary can easily partition line graphs and limit the set of potential senders. But if we consider the $k$-regular privacy subgraph (with larger values $k$), partitioning becomes more and more difficult, eventually leading to the set of potential senders being all nodes in the Bitcoin network.

Overall, in Dandelion++, the authors proposed a strategy that optimizes mid-way between the cases of known and unknown privacy subgraphs, eventually resulting in selecting a $4$-regular privacy subgraph.
However, since our analysis shows that it is relatively easy to learn the privacy subgraphs, Dandelion++'s analysis of known privacy subgraphs is more applicable, which we could comprehensively quantify and verify with our entropy-based results.

%% file: sections/Related.tex
\section{Related Work}
\label{Related}

Peer-to-peer anonymity designs have been proposed over the last decades for a variety of applications, such as web browsing~\cite{reiter1998crowds}, messaging~\cite{drac} or decentralized services~\cite{hoang2018empirical}. Blockchain-related applications have however recently been the main drivers for the design and implementation of peer-to-peer anonymous routing schemes, which are needed to protect the anonymity of transactions at the network layer. In this work, we focus on recent anonymity-enhancing schemes for cryptocurrency: Dandelion(++) and Lightning Network.
Dandelion++ is deployed in Monero~\cite{Monero} and is also under consideration for integration with Bitcoin, whereas LN is already deployed with more than 10K active nodes. We note however that our evaluation focuses on features of the anonymous routing scheme without relying on transaction information, which makes the framework applicable to any anonymous routing scheme, regardless of whether it is used for routing blockchain transactions or any other sort of payload. 

It must be noted that we are not the first to propose a probabilistic Bayesian approach to model and evaluate anonymity. The high-level idea has been used already in early works such as Vida \cite{danezis2009vida}, which aims to estimate anonymity for mixnets that are too large to evaluate analytically. Our contribution is thus that for the first time, we build a generic Bayesian framework tailored to study the anonymity of anonymous P2P routing and utilize it to evaluate concrete schemes currently proposed or deployed in cryptocurrency applications: Dandelion(++) and LN. Importantly, we propose novel techniques based on specific routing properties of these schemes to perform the analysis and show the framework's effectiveness.

Similarly, there are multiple researches that study LN's privacy and anonymity aspects. In ~\cite{romiti2020cross}, Romiti \emph{et al.} demonstrate a cross-layer attack where the LN nodes were mapped to their bitcoin addresses. In ~\cite{tikhomirov2020quantitative} Tikhomirov \emph{et al.} attempt to perform a primitive analysis of the payment paths that may be vulnerable to deanonymization attacks. They do so by selecting a set of influential adversaries (high-degree nodes) and finding the potential payment paths intercepted by the adversary nodes. But adversary nodes in the payment paths cannot ensure deanonymization as LN's design ensures that a node on the payment path cannot determine whether the previous hop is the actual originator or just the forwarder. Thus, a more nuanced analysis (as performed in our work) is required to analyze how and to what extent the adversary can deanonymize a transaction for which it acts as an intermediary by exploiting the anonymous routing scheme.

\color{black}
Another recent work~\cite{kumble2021lightning} by Kumble \textit{et al.} studies the impact of routing in LN on the anonymity provided to transactions by exploiting LN-specific transaction characteristics. The work considers the timelock value for each transaction to count the number of possible originators (or receivers) and thus estimates the size of the anonymity set (without distinguishing between higher or lower likelihood to be the originator). In comparison, our work calculates the individual probabilities for each possible originator.
Moreover, the focus of our contribution is to provide a generic and flexible framework that exploits routing constraints to evaluate anonymity. Additional transaction-related information that is available to the adversary can however be also incorporated. 
Thus, our framework can be easily extended to consider the transaction timelock values in addition to the routing algorithm details currently considered, highlighting its flexibility to incorporate various scheme-specific details while providing a common evaluation platform. 

Our analysis provides additional benefits as well. First, it is not influenced by mechanisms such as shadow routing \cite{shadow-routing}, which enables a client to add random timelock values making it difficult to guess the transaction recipient for any on-path transaction forwarder that just looks at the timelock values. Second, we consider transactions between each source-destination pair, which provides comprehensive coverage of all cases and a good estimate of the fraction of transactions captured by the adversary nodes.

Notably,~\cite{kumble2021lightning} considers all three LN client implementations (LND, eclair, and c-lightning), while our work focuses on the most widely used LND implementation ($>90\%$).  Considering all the client implementations for evaluation is ideal, but the analysis in ~\cite{kumble2021lightning} already shows that the additional minority implementations do not meaningfully impact the results: almost the same results were obtained with and without different implementation considerations.
Nonetheless, despite methodological differences to perform the evaluation, we note that the overall results and conclusions of both studies agree in the assessment that LN offers poor anonymity to its users.

\color{black}

Moreover, LN includes additional mechanisms such as $2$-of-$2$
multi-signature transactions (for channel construction), Hashed Time Lock Contracts (for payment management) \emph{etc.} \cite{poon2016bitcoin}. The exploitation of such mechanisms to deanonymize transactions has been studied in previous works~\cite{malavolta2017concurrency,green2017bolt,malavolta2018anonymous,rohrer2020counting,nisslmueller2020toward}. Our analysis makes abstraction of transaction data and instead relies exclusively on traffic data that can be passively collected. This makes our contribution generalizable for evaluating the anonymity provided by anonymous P2P routing schemes, whether they are used for routing Bitcoin transactions or simply messages.

%% file: sections/appendix.tex
\appendix
\section{Appendix}

\subsection{Reduction of Anonymity Set by a Sophisticated Adversary}
\label{subsec:reduce_anonymity_set}

\noindent \textbf{\textit{1) Can nuanced details of the privacy-subgraph help in reducing the anonymity?}}
In Dandelion++, each node has exactly two immediate successors (and can have more or less than two immediate predecessors) in the privacy-subgraph.
Dandelion++ further recommends that transactions received from a predecessor should be forwarded to a fixed successor.
However, in our analysis we do not incorporate these intricate details while computing entropy. This is because we believe that its hard for the adversary to know these internal predecessor--successor mappings of each node.

Thus in our analysis, for every node we consider that it sends transactions to \textit{any} one of its two successors, thereby having a more generic analysis, and recall that even without the knowledge of these mappings we obtained very low entropy values.
Notably, if some sophisticated adversary somehow obtains this predecessor--successor information, it can do a more precise analysis while estimating the originator of a received transaction.

\noindent \textbf{\textit{2) Can different originator probabilities reduce the anonymity?}}
In our current entropy computations, we consider a generic scenario, where we assume uniform priors for the originator probabilities ($P(B_i)$) for each benign node $i$. However, if a sophisticated adversary has additional knowledge (\eg knows the (average) frequency of transaction generation for each node $i$), it can utilize this extra information in the analysis by incorporating it into the priors. In such a case, the adversary would consider a different value of $P(B_i)$ for each node $i$ and perform a more informed analysis that further reduces entropy.

\noindent \textbf{Takeaway:} In spite of the fact that we do not incorporate the aforementioned details in our analysis (for all the three schemes), we obtained low anonymity values under various different settings. Our results are thus an upper bound, and a more sophisticated adversary can certainly perform a more precise analysis and deanonymize the transactions with even more confidence.

\subsection{Privacy subgraph learning}
\label{app:priv-learn}
In this section we now describe the approach in detail (mentioned in Sec.~\ref{subsec:grpah_learn}), that can be used by the adversary to obtain the privacy subgraph in Dandelion++. Subsequently, we also show the results obtained from simulations, and present some heuristics to learn privacy subgraph faster and with lower number of transactions.
We now describe the procedure to derive the privacy subgraph in simulations.
\begin{itemize}
    \item We construct a random 16-regular graph with each node having 8 outgoing neighbours. This is representative of the bitcoin graph (BG).
    \item Next, we derive a privacy subgraph (PSG) from the BG by selecting 2 outgoing connections randomly for each node as the successor nodes in the PSG.
    \item Since the adversary nodes already know their predecessor and successor connections, we add them to the derived PSG. 
    Adversary nodes would then send/forward the transaction via the remaining honest nodes. To ensure that at least one adversary node is connected to all honest nodes, every adversary node makes multiple connections to different honest nodes (instead of two).
    \item The adversary will then pick an honest node through which it will send multiple transactions and will record the nodes that diffuse them. We simulate the stem phase routing for these transactions. Since the forwarding probability assumed in Dandelion++ is 0.9, $\approx 10\%$ of all the transactions will be diffused by the successors of the honest node in the PSG. Thus, we analyze the number of diffusions by neighbours of the honest node in BG. The analysis reveals that the successors of the honest node in the PSG show significantly larger diffusions in comparison to other successors. We thus take the two nodes (out of the eight) with highest frequency as the actual successors of the node in PSG. We add these edges to the potential PSG derived by the adversary.
    We send 50 transactions per honest node before analyzing the distribution of diffusions per node.
    \item The previous step is repeated for all honest nodes and we obtain a PSG.
    \item We then compare the derived PSG with the actual PSG and calculate the accuracy.
\end{itemize}

Following the above steps we simulated the graph learning attack on a BG of 1000 nodes with $10\%$ adversary nodes. We observed that more than $90\%$ of the PSG was correctly constructed. Moreover, when we increased the number of transactions to be analysed per node from 50 to 100, the accuracy increased to $\approx 98.5$.

Additionally, in order to increase the accuracy of results, as well as to minimize the number of transactions to be analyzed per node, the adversary can employ extra analysis. The adversary can not only analyze the distribution of immediate successors of honest nodes, but also the successors of the immediate successors. This would allow the adversary to learn more edges in the graph with the same number of transactions.
It would also enable the adversary to further verify that the successor of the honest nodes the adversary identified are indeed correct. This is because if the successors are correctly identified then the successors corresponding to these successors would also show two nodes diffusing large number of transactions

Moreover, once the adversary has learned a large part of the PSG, it can then adopt additional strategies that may not require sending additional transactions, thereby minimizing the overall transactions required to learn the PSG. The adversary can do so by ruling out successors that are already part of the PSG and thus implicitly inferring the edges. This elimination will help adversary limit the set of potential successors and even in some cases will only be left with just the two successors that are actually part of the PSG.


